\documentclass[useAMS,usenatbib]{mn2e}

\title[Pc-scale morphology in faint high frequency
  peakers]{Parsec-scale morphology and spectral index distribution in faint high frequency peakers}
\author[M. Orienti \& D. Dallacasa]
  {M. Orienti$^{1,2}$\thanks{E-mail: orienti@ira.inaf.it},
D. Dallacasa$^{1,2}$\\
$^1$Dipartimento di Astronomia, Universit\'a di Bologna, via Ranzani
1, I-40127, Bologna, Italy\\
$^{2}$INAF - Istituto di Radioastronomia, Via P. Gobetti 101, I-40129 Bologna, Italy}

\date{Received \today; accepted ?}

\pagerange{\pageref{firstpage}--\pageref{lastpage}} \pubyear{2002}

\def\LaTeX{L\kern-.36em\raise.3ex\hbox{a}\kern-.15em
    T\kern-.1667em\lower.7ex\hbox{E}\kern-.125emX}

\begin{document}

\label{firstpage}

\maketitle

\begin{abstract}
We investigate the parsec-scale structure of 17 high frequency
peaking radio sources from the faint HFP sample. VLBA
observations were carried out at two adjacent frequencies, 8.4 and 15.3
GHz, both in the optically-thin part of the spectrum, to obtain the
spectral index information. We found that 64\% of the sources are
resolved into subcomponents, while 36\% are unresolved even at the highest
frequency. Among the resolved sources, 7 have a morphology and a
spectral index distribution typical of young radio sources, while in
other 4 sources, all optically associated with quasars, the radio properties
resemble those of the blazar population. 
The equipartition magnetic field of the single components 
are a few tens milliGauss, similar to the values found in
the hotspots of young sources with larger sizes. 
Such high magnetic fields cause severe radiative losses,
precluding the formation of extended lobe structures emitting at
centimeter wavelengths. The magnetic
fields derived in the various components of individual source are usually
very different, indicating a non self-similar source evolution, at
least during the very first stages of the source growth.

\end{abstract}


\section{Introduction}

Young radio sources can provide information
on the birth and evolution of the radio emission originating in the
central region of an active galactic nucleus (AGN). Powerful radio
emission L$_{\rm 1.4 GHz}>10^{24.5}$ W~Hz$^{-1}$ is found only in a
small fraction ($\sim$10\%)
of the AGN population. 
The population of extragalactic
radio sources is divided into several sub-classes each representing a
different stage in the individual source life. In the
evolutionary framework, the age of a radio source is directly related
to its linear size \citep[e.g.][]{fanti95,snellen00}. 
In this context, the empirical anticorrelation
found by \citet{odea97} 
between the intrinsic linear sizes, $LS$, and the spectral peak frequency
$\nu_{\rm p}$ indicates an evolutionary path linking the different
evolutionary stages: high frequency peakers (HFP, $\nu_{\rm p} >$ 5
GHz and $LS$ $\ll$100 pc), will evolve into gigahertz-peaked
spectrum (GPS, $\nu_{\rm p} \sim$ 1 GHz and $LS$$\leq$1 kpc), and then
into compact steep-spectrum (CSS, $\nu_{\rm p} \sim$ 100 MHz and $LS
\leq$15 kpc) objects, which are the progenitors of the classical
Fanaroff-Riley I/II radio galaxies (FRI/FRII, $LS$ up to a few Mpc,
Fanaroff \& Riley 1974). \\
The genuine youth of CSS/GPS sources was
proved by the determination of both the kinematic 
\citep{owsianik98a,owsianik98b,taylor00,tschager00,polatidis03,mo10a}
and radiative \citep{murgia03,mo07a,mo04} ages of a dozen of the most compact
objects ($LS \leq$ 100 pc), which turned out to be 10$^{3}$-10$^{4}$
years.\\
With their typical linear sizes up to a few tens of parsecs, the sub-class
of HFP should represent the very first stage in the radio source
evolution, with ages of a few hundred years. Their radio properties
have been derived by means of a detailed analysis of the flux density
variability \citep{tinti05,mo07}, pc-scale morphology \citep{mo06}
and polarization properties \citep{mo08} of the sources from
the ``bright'' HFP sample \citep{dd00}. From these studies it became
clear that the class of HFPs is made of two different populations. One
consists of genuinely young radio sources: they are
non-variable, with a double/triple structure resembling
a scaled-down version of the classical FRIIs, whose radio emission,
unpolarized, is dominated by hotspots/lobes. The other population is
made of blazar sources, characterized by high level of variability,
pc-scale core-jet structure and with significant polarized emission. Although
blazars usually display a flat radio spectrum, they may show a convex
peaked spectrum during particular phases of their variability,
i.e. when a self-absorbed knot in the jet dominates the radio emission. In
the case of the bright HFP sample, almost 60\% of the objects
have blazar-like properties.\\ 
A high percentage of blazar objects in
  samples of high-frequency peaking objects was found by
  \citet{torniainen05}, \citet{bolton06},
\citet{torniainen07}, and \citet{hancock09} by monitoring the
flux density variability of several HFP samples. However, among the
HFP/GPS sources optically associated with a galaxy the percentage of
genuinely young radio sources is larger than in objects hosted in
quasars  \citep[e.g.][]{hancock10,stanghellini05,tinti05}, 
indicating that projection 
and boosting effects play a role
in selecting objects in a high-activity state. \\
To investigate and compare the physical characteristics of HFPs
spanning a broader luminosity range (L$_{\rm 5 GHz}$ 
$\sim$ 10$^{24}$ - 10$^{28}$ W/Hz)
we performed analysis of the ``faint'' HFP sample
\citep{cstan09}. Flux density and spectral variability studies of the sources
in this sample were performed by means of multi-epoch VLA observations
carried out at various frequencies simultaneously \citep{mo10b}. From
this analysis it turned out that 42\% of the sources are non-variable,
while 21\% are highly variable with their radio spectrum becoming flat
at least in one observing epoch, as typical of the blazar population.
The remaining sources maintain the convex
peaked spectrum, but with some flux density variability. However, 37\%
of them possess a variability that can be explained in terms of a very
young radio source undergoing (adiabatic) expansion, suggesting
that a fraction of variable sources may still be genuinely young
objects.\\
In this paper we continue the analysis of the radio properties of
the sources from the ``faint'' HFP sample by
means of VLBA observations at 8.4 and 15.3 GHz. 
Among the 61 sources of
the sample we observed the 17 sources whose spectral peak occurred
at a frequency below 8.4 GHz.
The availability of two
observing frequencies in the optically-thin part of the spectrum 
allows us to derive the spectral index
distribution across the source structure to classify the nature of
each sub-component and thus of each source. In the classical
  picture, young radio
  sources are characterized by a two-sided structure dominated by
  lobes/hotspots, like the population of compact-symmetric objects
  (CSO). On the other hand, one-sided flat-spectrum core-jet structure
is typical of boosted sources.\\
The layout of the paper is the following: Section
2 describes the VLBA observations and the data reduction; Section 3
provides information on the radio morphology and spectral index
distribution for each source; a concise discussion and a brief summary
are reported in Sections 4 and 5.\\
Throughout this paper we assume 
$H_{0} = 71$ km s$^{-1}$ Mpc$^{-1}$, $\Omega_{\rm M} = 0.27$,
$\Omega_{\Lambda} = 0.73$, in a flat Universe. The spectral index
$\alpha$ is
defined as $S$($\nu$) $\propto \nu^{- \alpha}$. \\

\section{VLBA observations and data reduction}

We performed VLBA observations at 8.4 and 15.3 GHz of 17 sources from
the faint HFP sample \citep{cstan09}. The targets were chosen on the
basis of their spectral peak occurring at a frequency below 8.4 GHz, in order 
to study the
optically-thin part of the spectrum. VLBA observations were carried
out between February and March 2010 (Table \ref{log}), in single
polarization with an aggregate bit-rate of 256 Mbps. The correlation
was performed at the VLBA correlator in Socorro.\\
Each source was
observed for about 30 min at 8.4 GHz and 1 hr at 15.3 GHz, spread into
7 to 9 short scans of about 3 minutes each, switching
between frequencies and sources in order to improve the coverage of
the {\it uv}-plane. The strong sources DA\,193, and 4C\,39.25
were used as fringe finders and bandpass calibrators. 
Phase referencing 
was performed for the 8 targets fainter
than 35 mJy. Due to their proximity to the phase
calibrators used for other targets, we could observe three additional
sources (J0804+5431, J1319+4851, and J1436+4820) in
phase-referencing mode in order to determine their
accurate position.\\ 
The data handling was carried out by
means of the NRAO \texttt{AIPS} package. A priori amplitude calibration was
derived using measurements of the system temperature and the antenna
gains. The errors on the absolute flux density scale were estimated
by comparing the data of either DA\,193 or 4C\,39.25 with the values
reported in the VLA/VLBA database. They resulted
to be within 10\% ($\sigma_{\rm c}$). \\
Final images were produced after a number of phase self-calibration
iterations. Amplitude self-calibration was applied at the end of
the process only for the brightest sources, 
in order to remove residual systematic errors, and using a
solution interval longer than the scan-length. The flux densities of
our targets are
generally in agreement with those measured with the VLA. The
  rms noise level on the image plane is between 0.05 and 0.25
  mJy/beam. Therefore, the main uncertainty in the flux densities
  generally comes
  from the amplitude calibration errors.\\

\begin{table}
\caption{Log of the VLBA observations}
\begin{center}
\begin{tabular}{cccc}
\hline
Code& Obs. date& Obs. time&Info\\
\hline
BD147A&15 Feb 2010& 5.5 hr& No SC\\
BD147B&20 Feb 2010& 7.0 hr& - \\ 
BD147C&04 Mar 2010& 7.0 hr& - \\
BD147D&01 Mar 2010& 4.5 hr& No SC\\
\hline
\end{tabular}
\end{center}
\label{log}
\end{table}

\section{Results}

Of the 17 sources 
with VLBA observations, 6 are resolved at both frequencies, while
5 sources are resolved at the higher frequency only. The remaining 6
sources are unresolved even with the high resolution provided by
the 15.3 GHz VLBA observations (FWHM $\sim$ 1 mas). \\
Throughout the paper we consider
``marginally resolved'' (MR) those sources whose largest angular size (LAS)
detected is between 0.5 and 1 times the beam size at 15.3 GHz, and we term
"Unresolved'' (Un) all the sources whose LAS is 
smaller than half the beam size at both frequencies. \\
The total flux
density of each source is reported in Table \ref{sample} 
and was measured by means of TVSTAT which performs an
aperture integration over a selected region on the image plane. 
For the resolved sources we also derived the flux density and the
deconvolved angular size of each source component by means of the AIPS
task JMFIT which performs a Gaussian fit to the source components on
the image plane. The formal uncertainty on the deconvolved angular size
  obtained from the fit is $\leq$0.1 mas. 
As expected, the sum of the flux density from each component
agrees with the total flux density measured on the whole source
structure. Source components are referred to as North (N), South (S),
East (E), West (W), and Central (C).
Observational parameters of the source components are
reported in Table \ref{tab_comp}.\\
The uncertainty on the flux density arises from both the calibration
error $\sigma_{\rm c}$ (see Section
2), and the measurement error $\sigma_{\rm m}$. 
The latter represents the off-source noise level, rms,
measured on the image plane and it is related to the source size
$\theta_{\rm obs}$ normalized by the beam size $\theta_{\rm beam}$ as
$\sigma_{\rm m}= {\rm rms} \times (\theta_{\rm obs}/ \theta_{\rm
  beam})^{1/2}$. The
flux density errors $\sigma_{\rm S}$ reported in Tables \ref{sample} and
\ref{tab_comp} take into consideration both uncertainties, and they
correspond to $\sigma_{\rm S} = \sqrt{ \sigma_{\rm c}^{2} + \sigma_{\rm
m}^{2}}$. Errors on the spectral index $\alpha_{\rm 8.4}^{\rm 15}$ 
have been computed
assuming the error propagation theory. In general the uncertainty on
the overall component spectral index is $\sim$0.1-0.2 in the most
compact and brightest components, while it becomes larger in the
weakest ones (Tables \ref{sample} and \ref{tab_comp}). \\
For those sources observed with the phase-referencing technique we
could derive the accurate absolute position, reported in Table
\ref{astrometry}, not available
previously. 
In fact, the majority of the targets lack previous
mas-resolution observations, and the coordinate information was 
obtained from
the FIRST survey \citep{becker95}. In earlier multifrequency VLA
observations, the snapshots were very short, and for most sources the
secondary calibrator was often too far away to provide accurate
information on the source position. The uncertainty on the source
position derived from these VLBA data
is about 0.5 mas. Such uncertainties have been estimated for the sources
observed in phase referencing at both frequencies, by comparing the
positions derived at both 8.4 and 15.3 GHz.\\

\subsection{Source images}

Full-resolution images of the 6 sources with resolved structure at
both frequencies are presented in Fig. \ref{cso}. In
Fig. \ref{mr} we present the full-resolution images at 15.3 GHz of the
sources which are marginally resolved at this frequency only. \\
For the sources with resolved morphology at both frequencies, 
to produce spectral index images in addition to the
full-resolution images, we produced also low-resolution images at both
frequencies using the same {\it uv}-range between 11 M$\lambda$ and
240 M$\lambda$. Furthermore, the images have been produced at both
frequencies with the same image sampling, natural grid weighting and,
in the case of the 15.3 GHz,
by forcing the beam major
and minor axes, and position angle to be equal to those of the 8.4-GHz
image.\\
The spectral index images were produced by combining the
low-resolution images at both frequencies by means of the AIPS
  task COMB. Blanking was done clipping the pixels of the input images
with values below three times the rms measured on the off-source image
plane at each frequency.
Image registration was
performed by comparing the position of the source components and  
applying an image shift using the \texttt{AIPS} task LGEOM, when necessary. 
The greyscale spectral index images superimposed on
the 15.3 GHz contours are presented in Fig. \ref{spix}. Despite the
particular care used to produce these images, some gradients in the
direction transverse to the source axis are present in a few sources, 
likely due to the
poor {\it uv}-coverage of these short observations. This is
particularly critical in the case of J0943+5113 where the artificial
gradients do not allow us to reliably measure the spectral index across
the Western component.   \\

\begin{table*}
\caption{The VLBA sample. Column 1: source name (J2000); Col. 2:
  optical identification; Q=quasar, G=galaxy, EF=empty field; 
  Col. 3: redshift; a ``p'' indicates a photometric redshift from
  \citet{mo10b}; Cols. 4, 5: VLA flux density at 8.4 and 15.3 GHz
  respectively, from \citet{cstan09}; Cols. 6, 7: VLBA flux density at
  8.4 and 15.3 GHz, respectively; Col. 8: luminosity at 1.4 GHz, computed using the flux density at 1.4 GHz reported in \citet{cstan09}; 
Col. 9: VLBA spectral index computed between 8.4
  and 15.3 GHz;
  Col. 10: VLBA structure: Res=resolved; MR=marginally
  resolved; Un=unresolved; Col. 11: source classification: CSO=
  genuine CSO; CSO?=CSO
  candidate; BL=blazar-like object; ?=not enough information to
  reliably classify the source nature.} 
\begin{center}
\begin{tabular}{lcccccccccc}
\hline
Source&ID&z&S$_{\rm 8.4}^{\rm VLA}$&S$_{\rm 15.3}^{\rm VLA}$&S$_{\rm
  8.4}^{\rm VLBA}$&S$_{\rm 15.3}^{\rm VLBA}$&$L_{\rm 1.4 GHz}$&$\alpha_{\rm 8.4}^{\rm
  15}$&Morph&Class\\
 & & &mJy&mJy&mJy&mJy&W/Hz& & &\\
(1)&(2)&(3)&(4)&(5)&(6)&(7)&(8)&(9)&(10)&(11)\\
\hline
&&&&&&&&&&\\
J0736+4744& Q&    -   &  47&  30&  45$\pm$5&  33$\pm$3&
26.86 & 0.5$\pm$0.2& MR&?\\
J0804+5431& G&  0.22p &  67&  49&  53$\pm$5&  41$\pm$4&
25.48 & 0.4$\pm$0.2& MR&?\\
J0819+3823& Q&    -   &  91&  44&  86$\pm$8&  46$\pm$5&
27.11 & 1.1$\pm$0.2& Un&?\\
J0905+3742&EF&    -   &  68&  39&  65$\pm$6&  40$\pm$4&
26.99 & 0.8$\pm$0.2& Res&CSO?\\
J0943+5113& G&  0.42p &  64&  26&  59$\pm$6&  23$\pm$2&
26.10 & 1.4$\pm$0.2& Res&CSO?\\
J0951+3451& G&  0.29p &  55&  38&  55$\pm$6&  52$\pm$5&
25.66 & 0.1$\pm$0.2& Res&CSO\\
J0955+3335& Q&  2.491 & 105&  70&  52$\pm$5&  40$\pm$4&
28.09 & 0.4$\pm$0.2& Un&BL\\
J1002+5701&EF&    -   &  71&  20&  53$\pm$5&  13$\pm$2&
27.02 & 2.3$\pm$0.3& MR&CSO?\\
J1008+2533& Q&  1.960 &  96&  74&  88$\pm$9& 128$\pm$13&
27.83 &-0.6$\pm$0.2& Res&BL\\
J1107+3421&EF&    -   &  51&  32&  32$\pm$3&  20$\pm$2&
26.87 & 1.0$\pm$0.2& Res&CSO?\\
J1135+3624&EF&    -   &  41&  22&  41$\pm$4&  18$\pm$2&
26.78 & 1.4$\pm$0.2& Res&CSO?\\
J1241+3844& Q&    -   &  21&  17&  15$\pm$2&  16$\pm$2&
26.49 & 0.1$\pm$0.3& Un&BL\\
J1309+4047& Q&  2.910 &  95&  59&  85$\pm$9&  37$\pm$4& 28.22
& 1.4$\pm$0.3& MR&CSO?\\   
J1319+4851& Q&    -   &  37&  27&  33$\pm$3&  37$\pm$4&
26.73 &-0.2$\pm$0.2& MR&BL\\
J1420+2704& Q&    -   &  51&  37&  61$\pm$6&  44$\pm$4&
26.88 & 0.5$\pm$0.2& Un&?\\
J1436+4820&EF&    -   &  62&  40&  49$\pm$5&  29$\pm$3&
26.95 & 0.9$\pm$0.2& Un&?\\
J1613+4223& Q&    -   & 122&  44& 100$\pm$10&  31$\pm$3&
27.25 & 2.0$\pm$0.2& Un&?\\    
&&&&&&&&&&\\
\hline
\end{tabular}
\end{center} 
\label{sample}
\end{table*}

\begin{table*}
\caption{Parameters of the components of resolved sources. Column 1:
  source name; Col. 2: source component labelled as in
  Fig. \ref{cso}; Cols. 3, 4: flux density at 8.4 and 15.3 GHz,
  respectively; Cols. 5, 6: deconvolved major and minor axis, respectively;
  Col. 7: position angle of the major axis; Col. 8: spectral index
  between 8.4 and 15.3 GHz computed considering the total flux
    density of the whole component measured from the full-resolution
    images at 8.4 and 15.3 GHz (see Section 3.2); 
Cols. 9 and 10: angular and linear separation
from the brightest component; Col. 11: equipartition magnetic field (see Section 4.3).
 When neither a spectroscopic nor
photometric redshift is available, we compute the linear
separation assuming $z=1.0$.}
\begin{center}
\begin{tabular}{ccccccccccc}
\hline
Source&Comp&S$_{\rm 8.4}$&S$_{\rm 15.3}$&$\theta_{\rm max}$&$\theta_{\rm
min}$&PA&$\alpha$&Sep.&Lin.&$H_{\rm eq}$\\
 & &mJy&mJy&mas&mas& & &mas&pc&mG\\
(1)&(2)&(3)&(4)&(5)&(6)&(7)&(8)&(9)&(10)&(11)\\
\hline
&&&&&&&&&\\
J0905+3742& E&   51$\pm$5&  30$\pm$3&  
  0.722$_{-0.005}^{+0.005}$&  -   & 133$\pm$1 &
0.9$\pm$0.2&    &    &$>$60\\ 
          & W&   14$\pm$2&  10$\pm$2& 
  1.692$_{-0.029}^{+0.030}$&   0.790$_{-0.068}^{+0.063}$&
  74$\pm$2&
0.6$\pm$0.4& 1.7$\pm$0.1& 13.7$\pm$0.8&10\\
J0943+5113& E&   17$\pm$2&  10$\pm$2&   $<$0.4&   $<$0.2&  - &
0.9$\pm$0.4&    &    &$>$27\\
          & W&   43$\pm$4&  12$\pm$2& 0.769$\pm$0.005&
  0.521$_{-0.169}^{+0.119}$&  73$\pm$20&
2.1$\pm$0.3& 3.9$\pm$0.2& 21.5$\pm$1.1&13\\
J0951+3451& E&   18$\pm$2&  13$\pm$2& 
  0.331$_{-0.021}^{+0.022}$&   $<$0.1 & 8$\pm$4&
0.5$\pm$0.3& 2.3$\pm$0.1& 9.9$\pm$0.4 &$>$40\\
          & C&   33$\pm$3&  33$\pm$3& 
  0.355$_{-0.003}^{+0.004}$& 0.203$_{-0.019}^{+0.017}$&
  58$\pm$2&
0.0$\pm$0.2&    &    &35\\
          & W&    4$\pm$1&   2$\pm$1&
  1.452$_{-0.118}^{+0.113}$& 0.564$_{-0.111}^{+0.086}$& - &
1.1$\pm$0.9& 2.5$\pm$0.2 & 10.8$\pm$0.9 &7\\
J1008+2533& W&   58$\pm$6& 115$\pm$12& 0.120$\pm$0.004&
  0.082$\pm$0.012&
54$\pm$8&-1.1$\pm$0.2& 1.0$\pm$0.1& 8.5$\pm$0.8 &210\\
          & C&  -  &   3$\pm$1&    - &    - &  - &  - &    &    & \\
          & E&   30$\pm$3&  10$\pm$1& 0.776$\pm$0.030& 0.257$_{-0.105}^{+0.074}$& 106&
1.8$\pm$0.2& 1.6$\pm$0.1& 13.6$\pm$0.8 &41\\   
J1107+3421& E&   21$\pm$2&  13$\pm$1& 0.770$\pm$0.013 &
  0.225$_{-0.019}^{+0.016}$& 7$\pm$1&
0.8$\pm$0.2&    &    &$>$46\\
          & W&   11$\pm$1&   6$\pm$1&
  0.323$_{-0.136}^{+0.097}$&  0.304$_{-0.141}^{+0.106}$ & 130$\pm$40 &
1.0$\pm$0.3& 1.2$\pm$0.2 & 9.6$\pm$1.6 & 12\\  
J1135+3624& W&   28$\pm$3&  11$\pm$2& 0.252$\pm$0.032 & 0.138$_{-0.029}^{+0.134}$ &  - &
1.5$\pm$0.3&    &    &60\\
          & E&   13$\pm$1&   6$\pm$1&
  1.180$_{-0.161}^{+0.150}$ & 0.886$_{-0.173}^{+0.146}$ &
31$\pm$16 &
1.3$\pm$0.3& 1.3$\pm$0.2 & 10.4$\pm$1.6 &11\\

&&&&&&&&&\\
\hline
\end{tabular}
\end{center} 
\label{tab_comp}
\end{table*}

\begin{figure*}
\begin{center}
\includegraphics{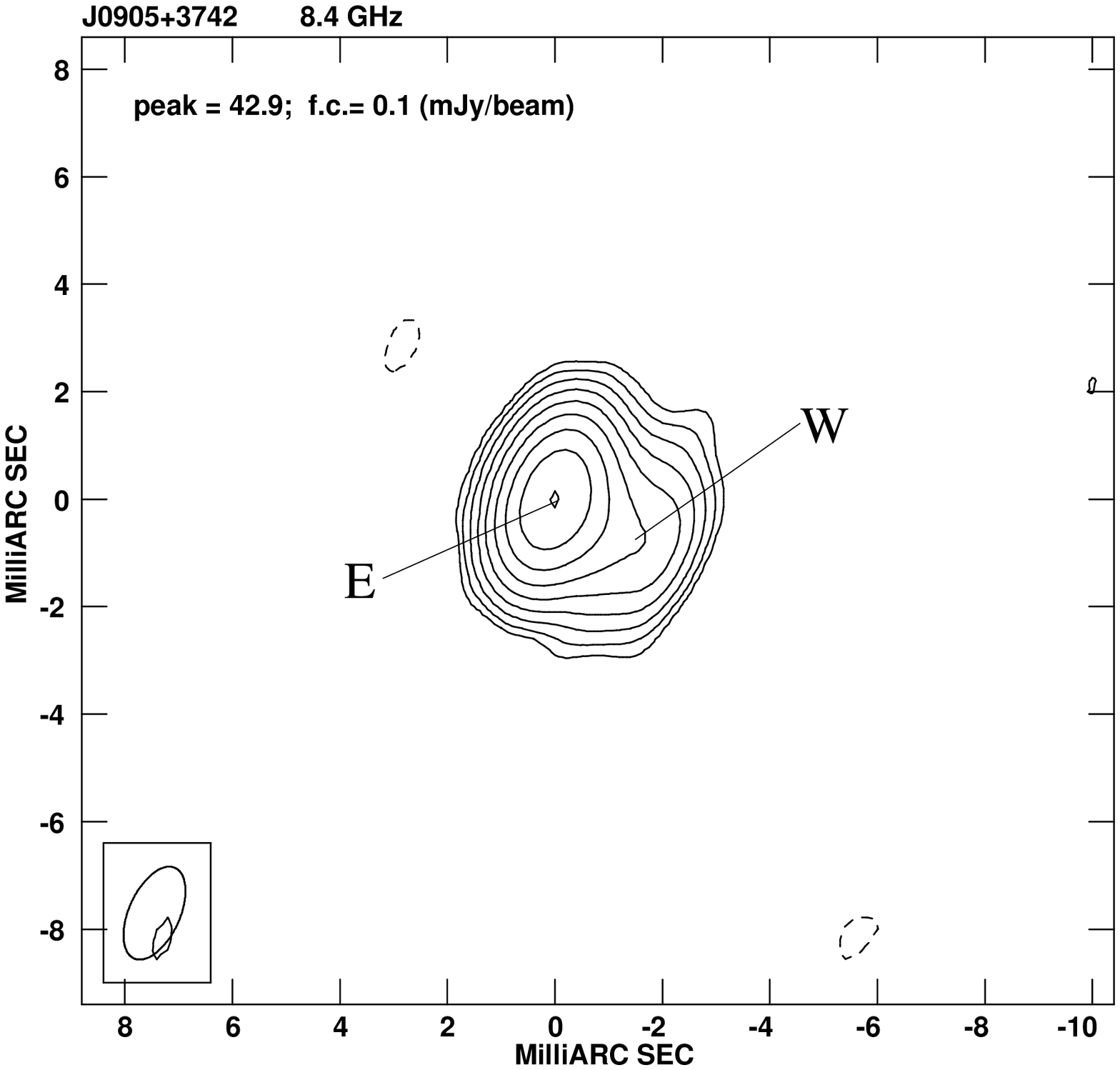}
\includegraphics{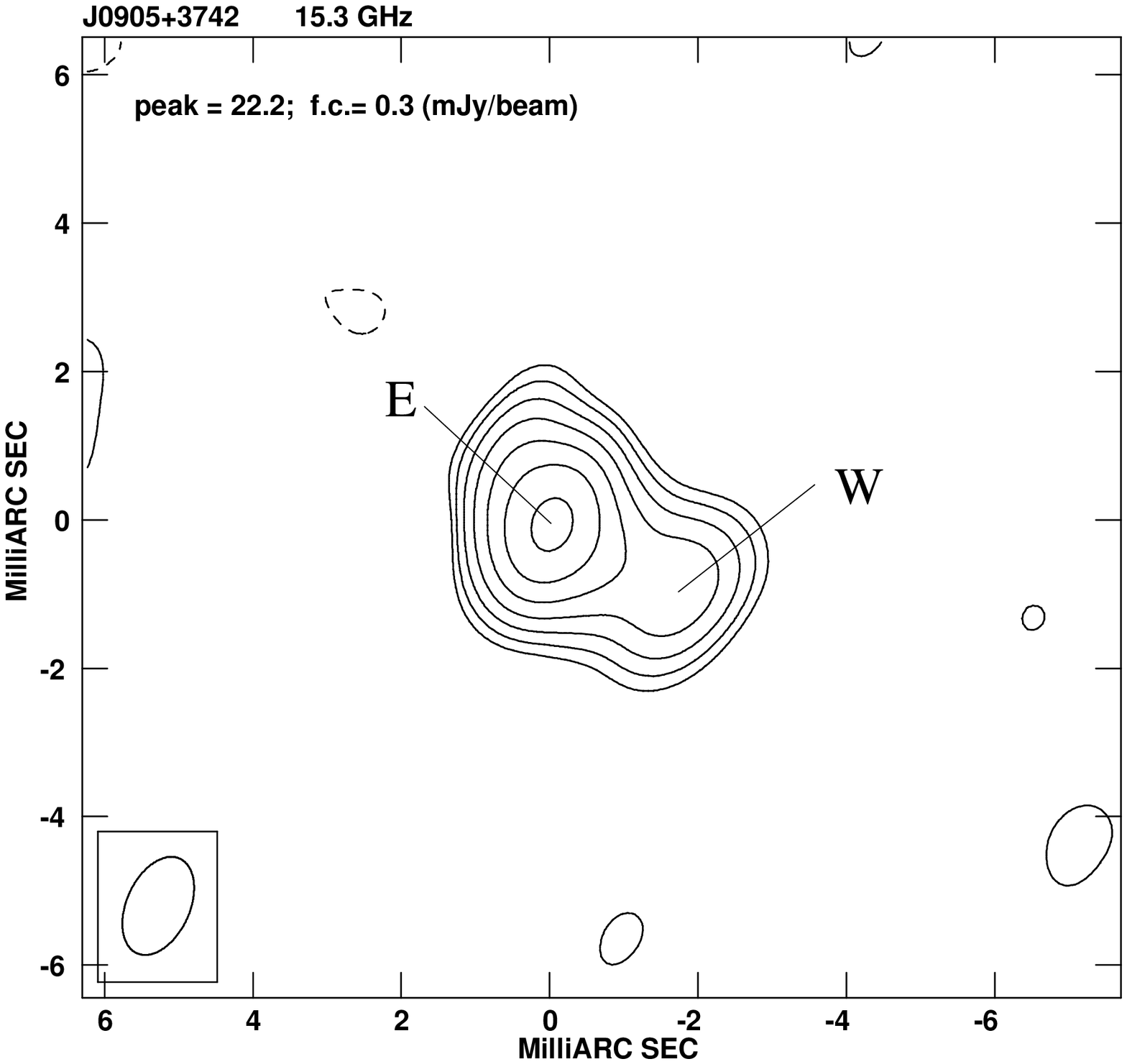}
\includegraphics{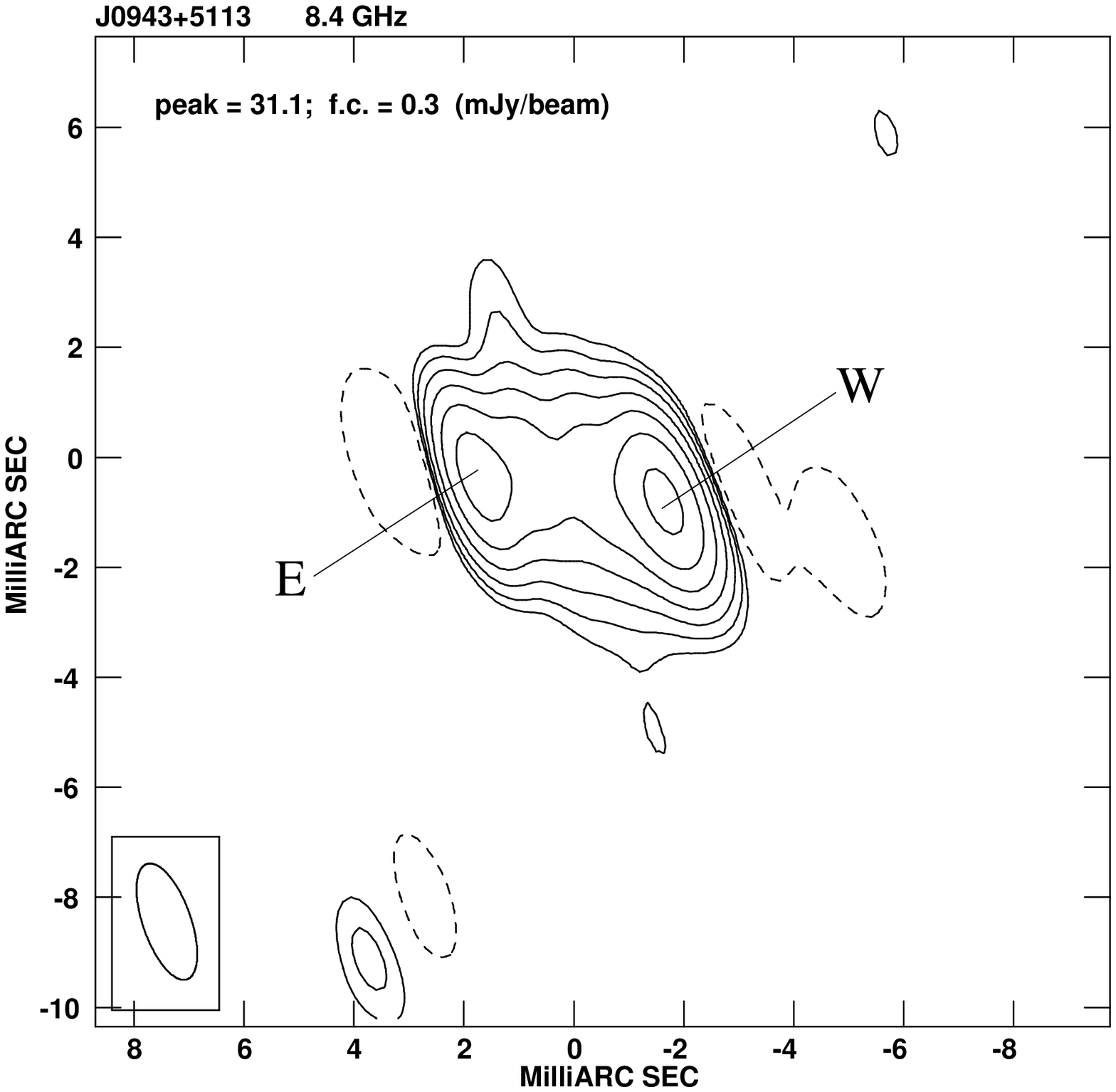}
\includegraphics{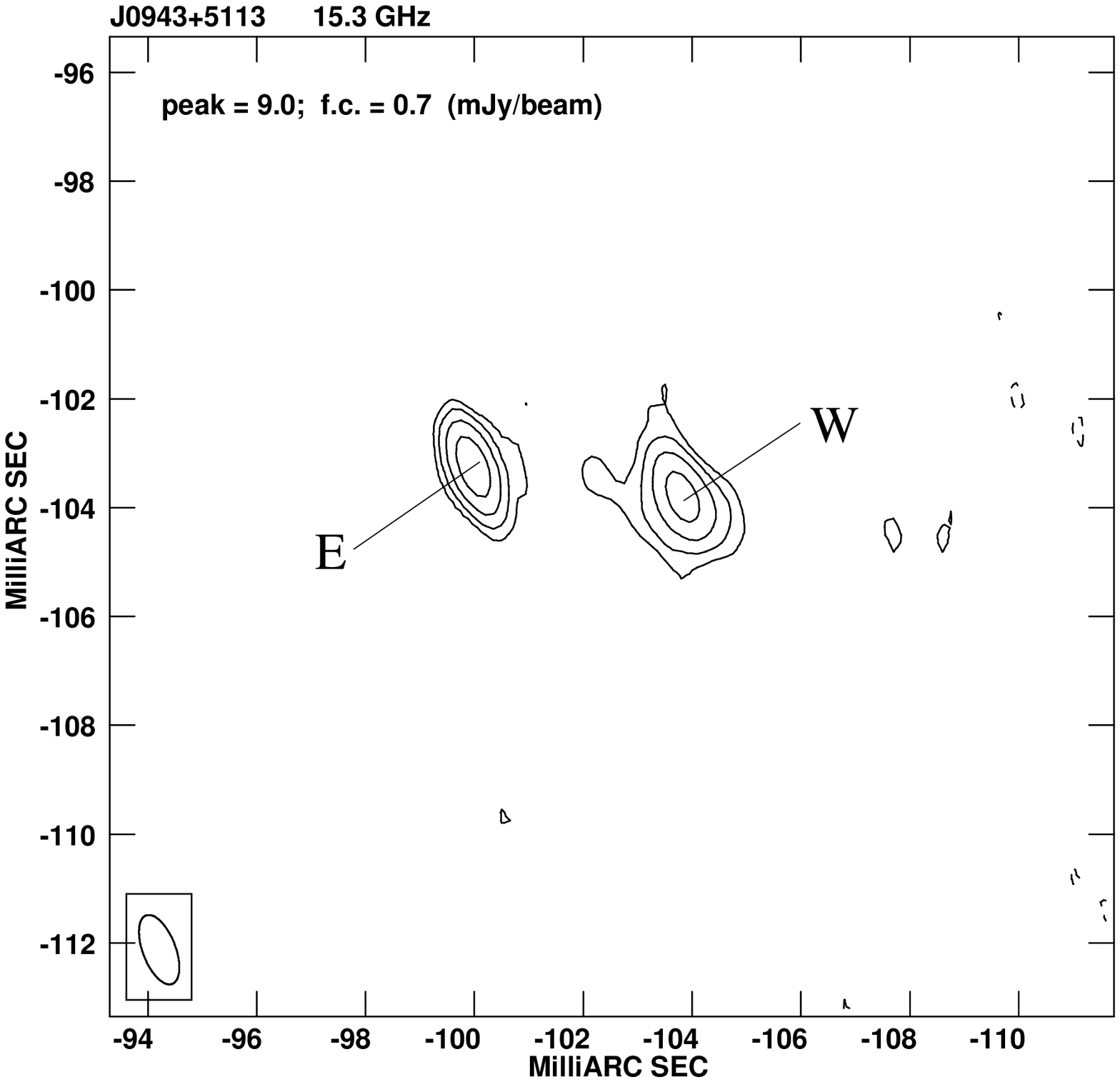}
\includegraphics{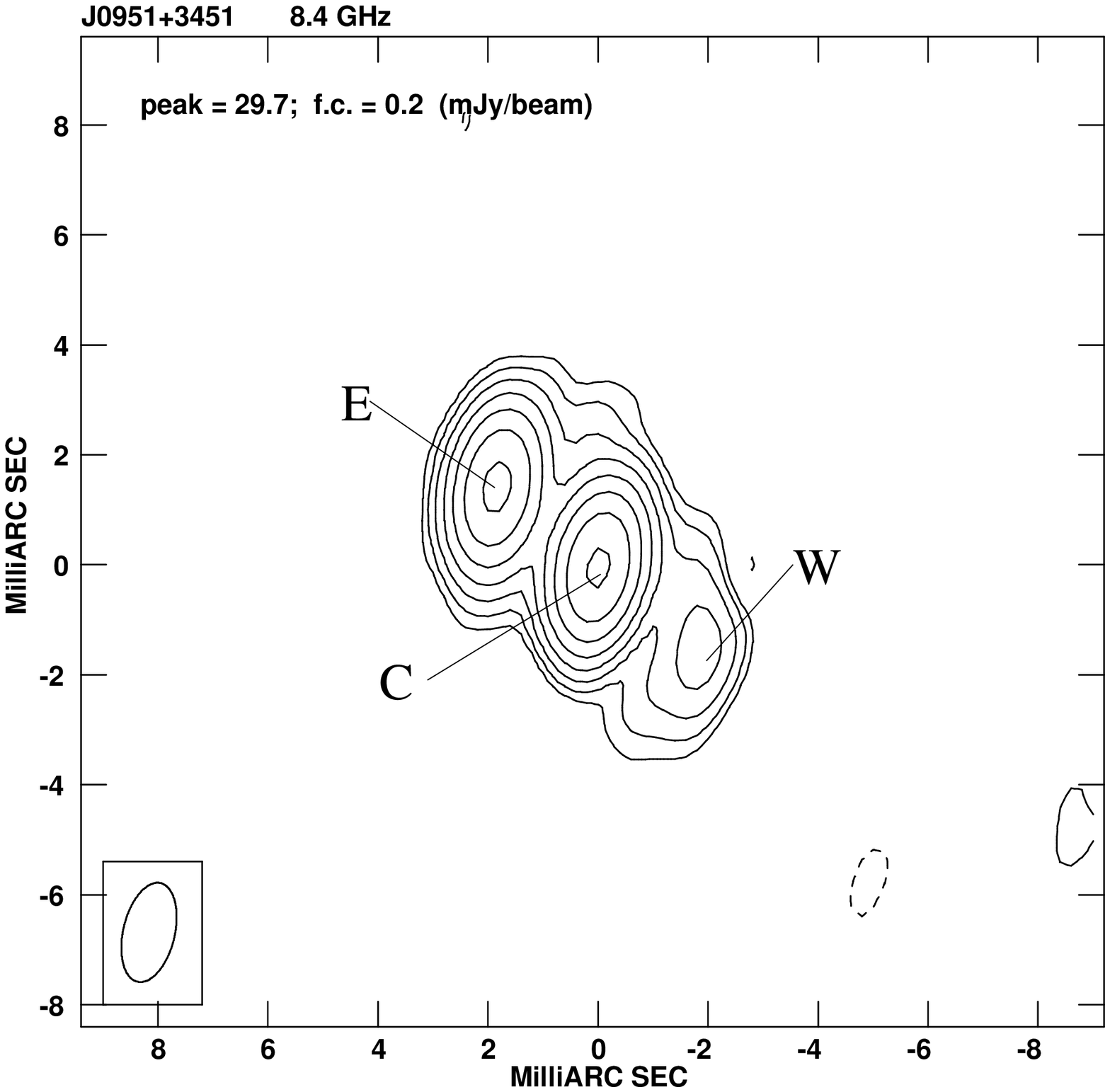}
\includegraphics{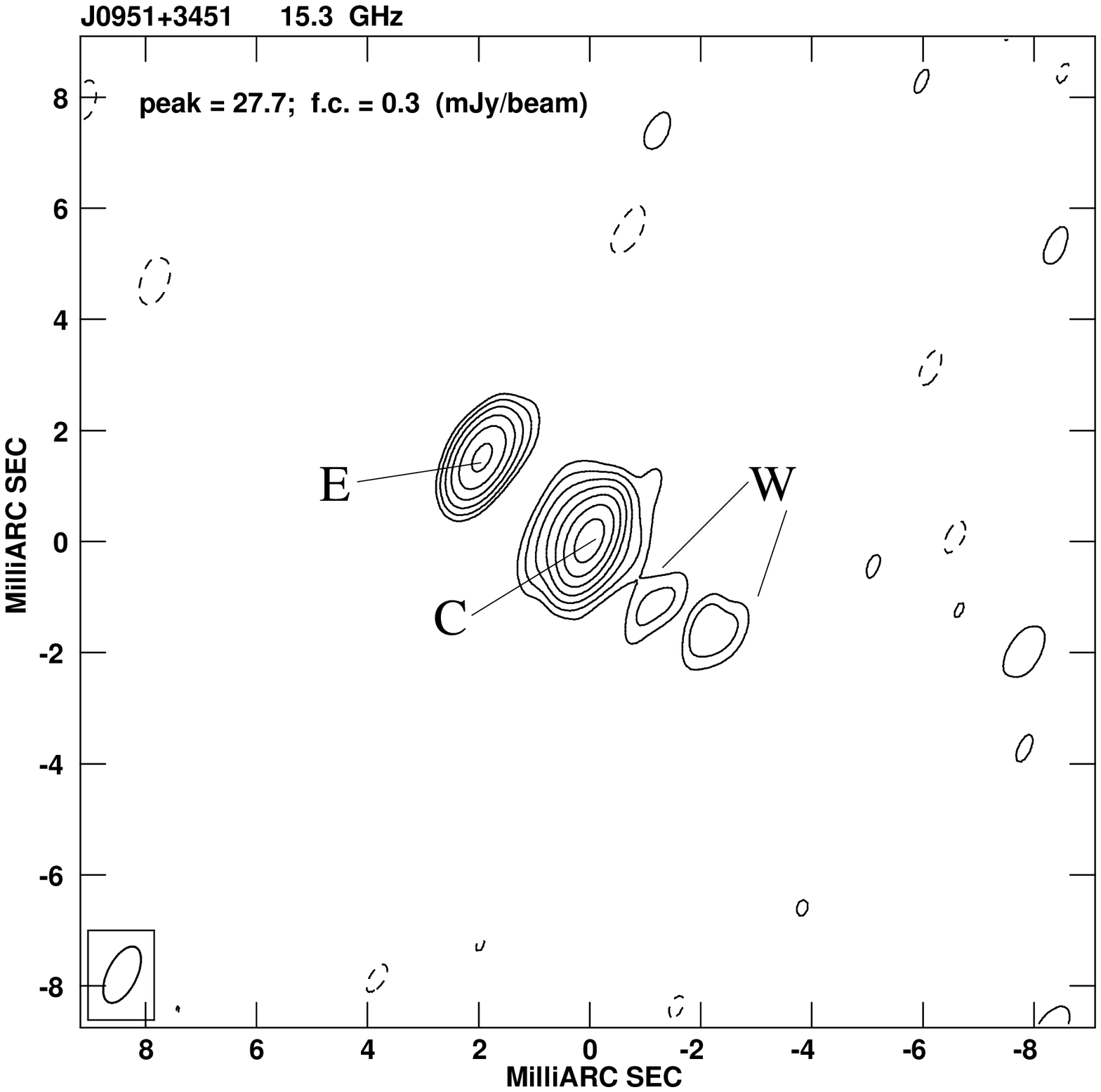}
\vspace{21.5cm}
\caption{VLBA images at 8.4 ({\it left}) and 15.3 GHz ({\it right}) of
  the sources with a resolved structure at both frequencies. For each
  image we provide the observing frequency, peak flux density
  (mJy/beam), the first contour intensity which corresponds to three
  times the off-source noise level on the image plane. 
Contours increase by a
  factor of 2. The restoring beam is plotted on the bottom left
  corner. Positions are relative to those used in the correlator.}
\label{cso}
\end{center}
\end{figure*}

\addtocounter{figure}{-1}
\begin{figure*}
\begin{center}
\includegraphics{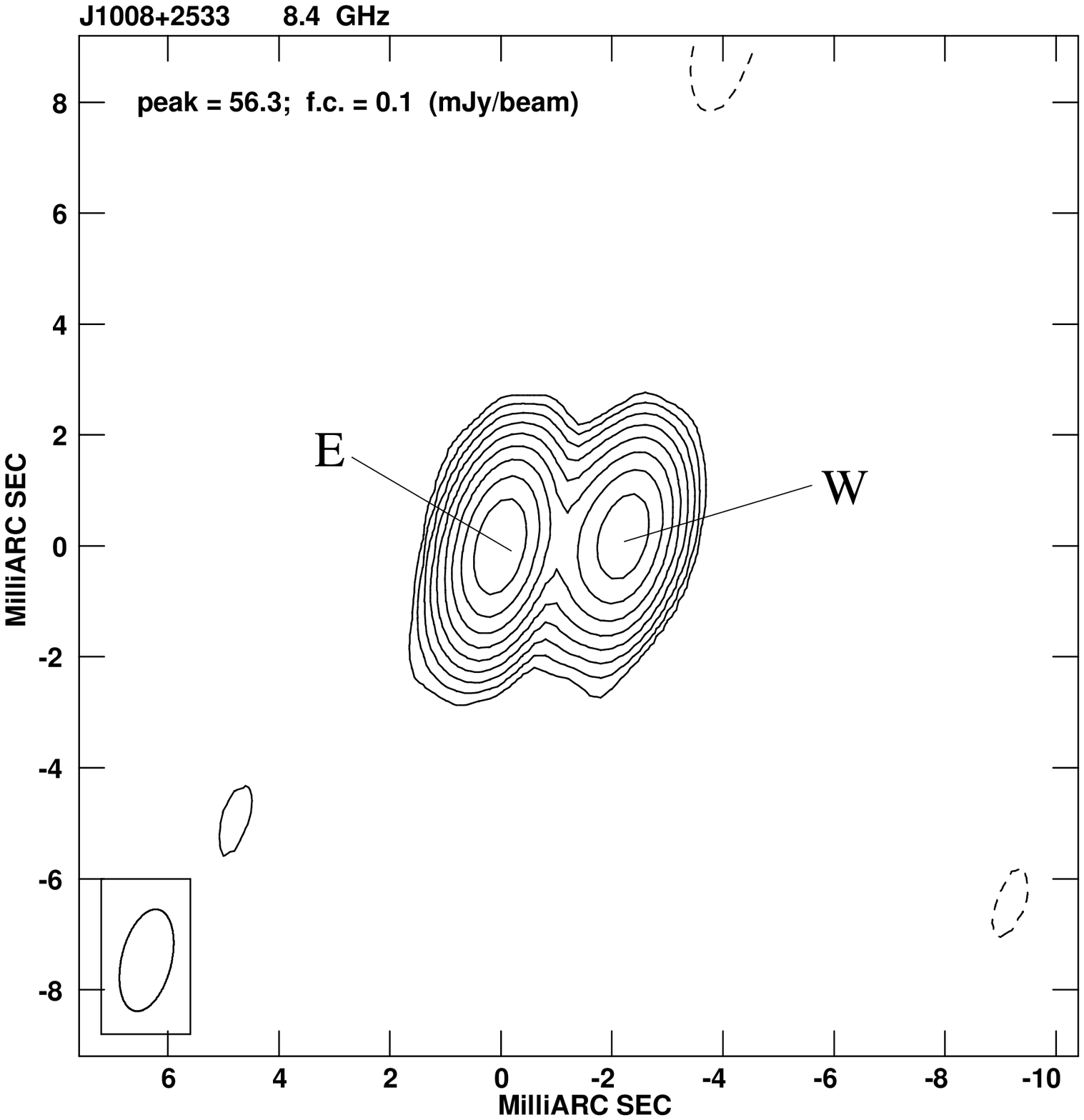}
\includegraphics{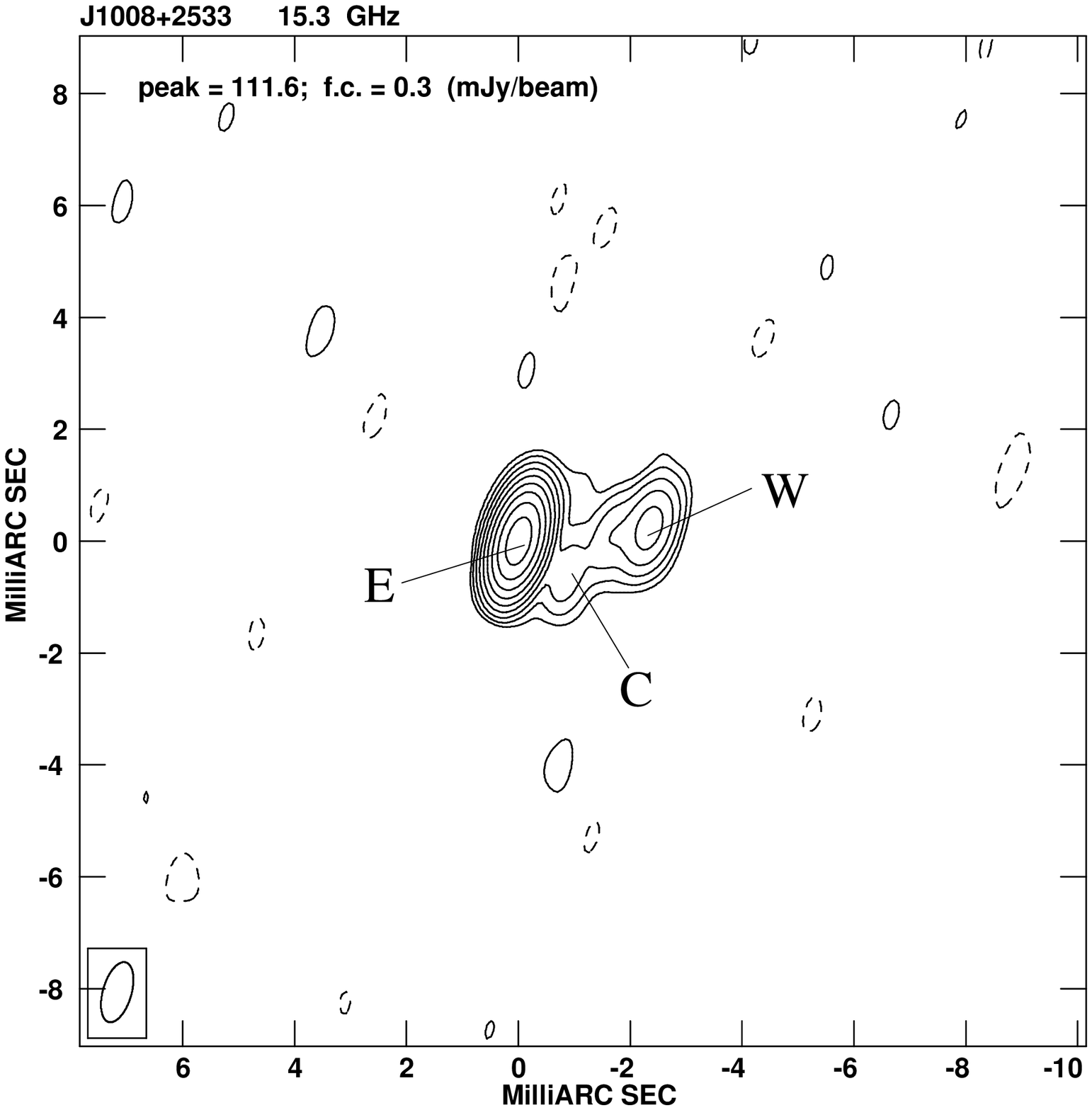}
\includegraphics{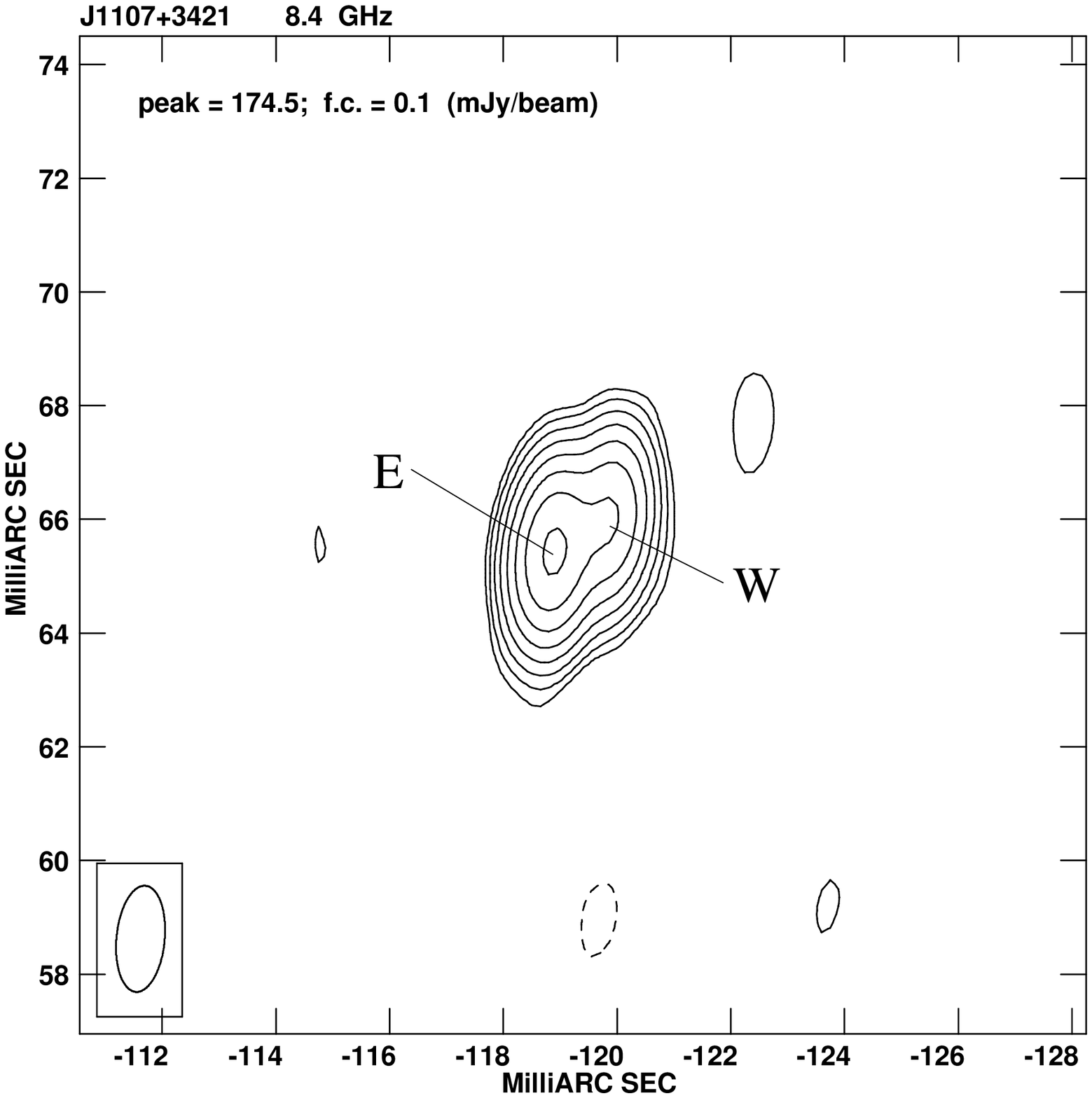}
\includegraphics{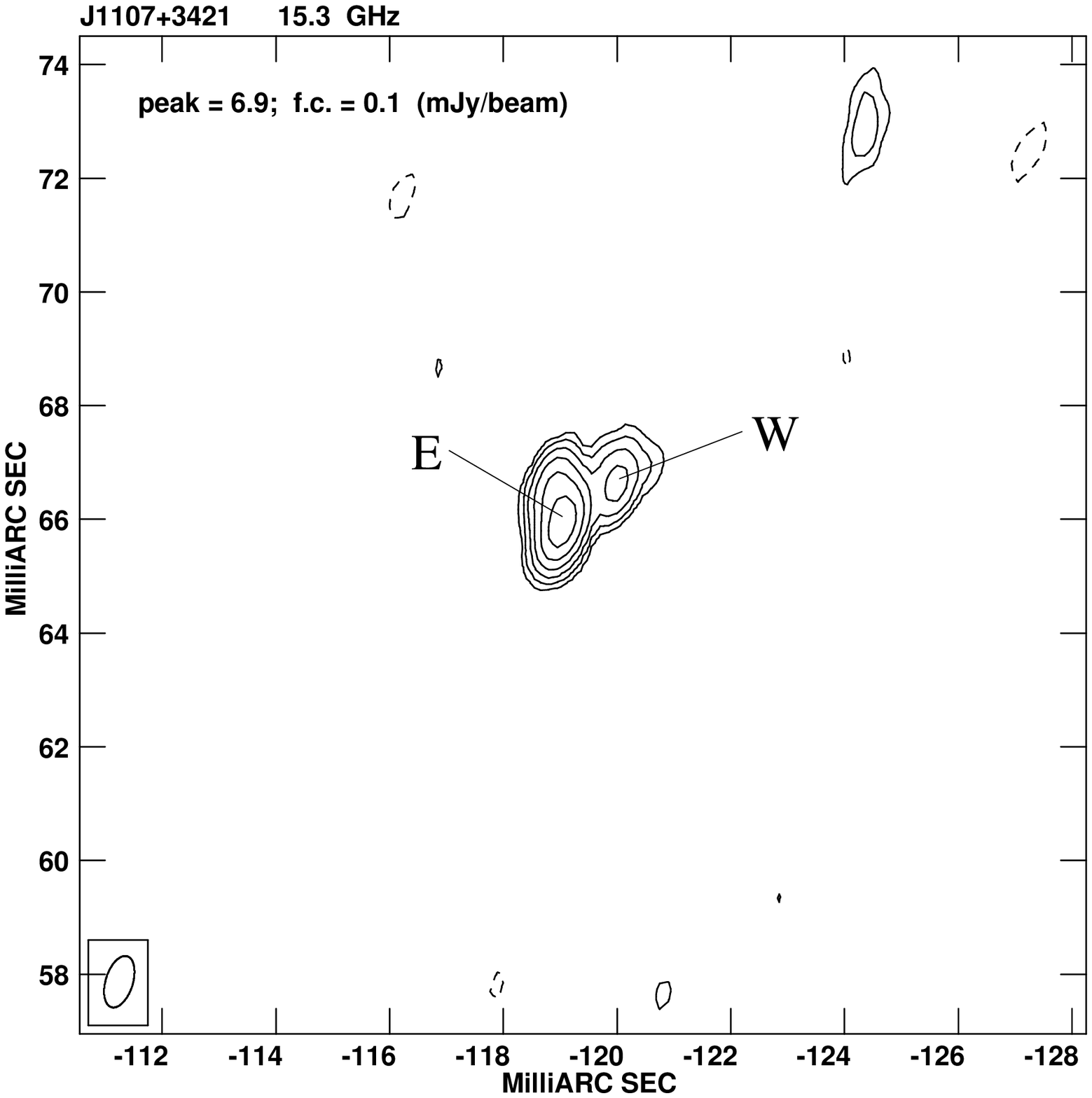}
\includegraphics{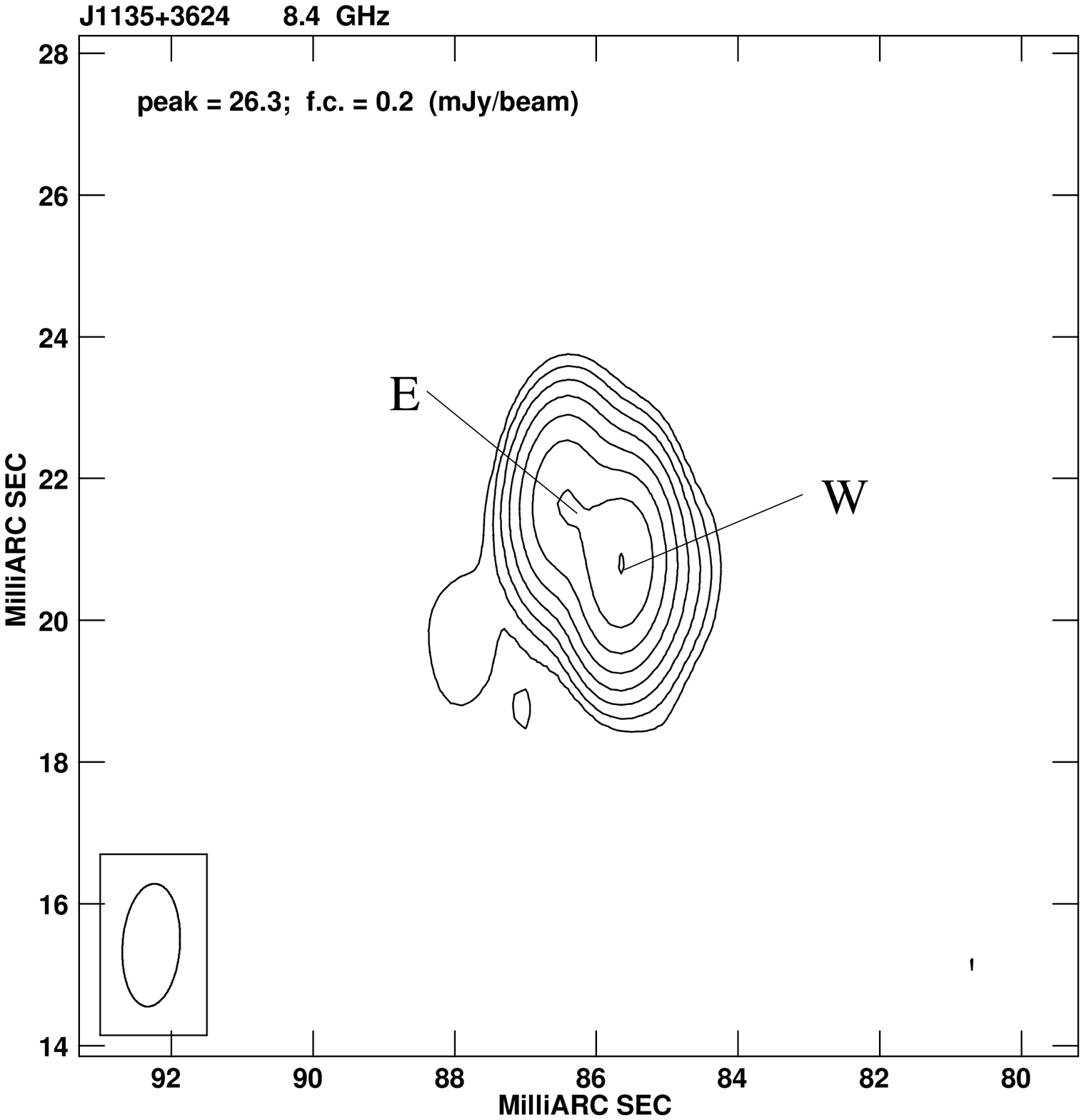}
\includegraphics{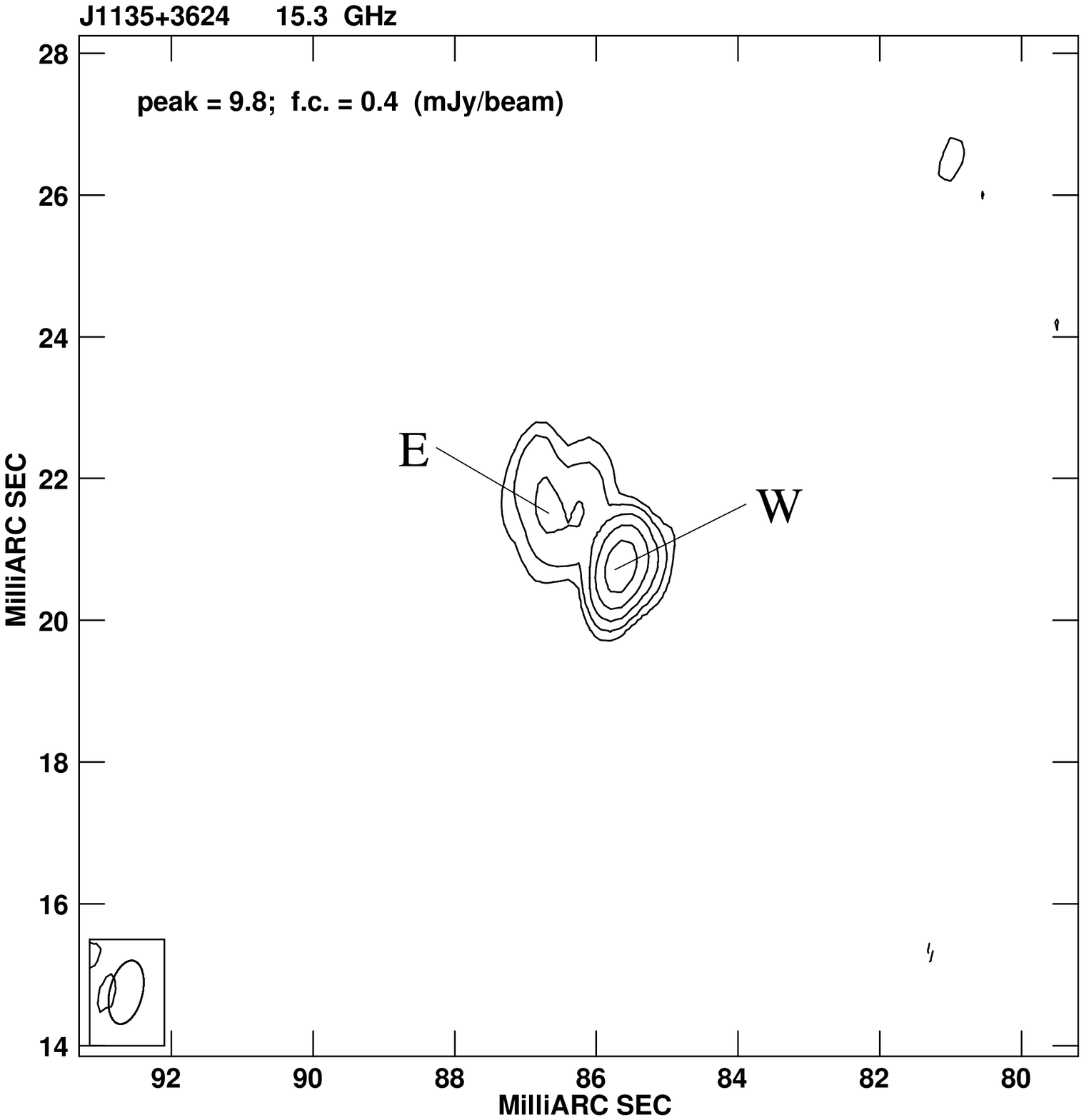}
\vspace{21.5cm}
\caption{Continued.}
\end{center}
\end{figure*}

\begin{figure*}
\begin{center}
\includegraphics{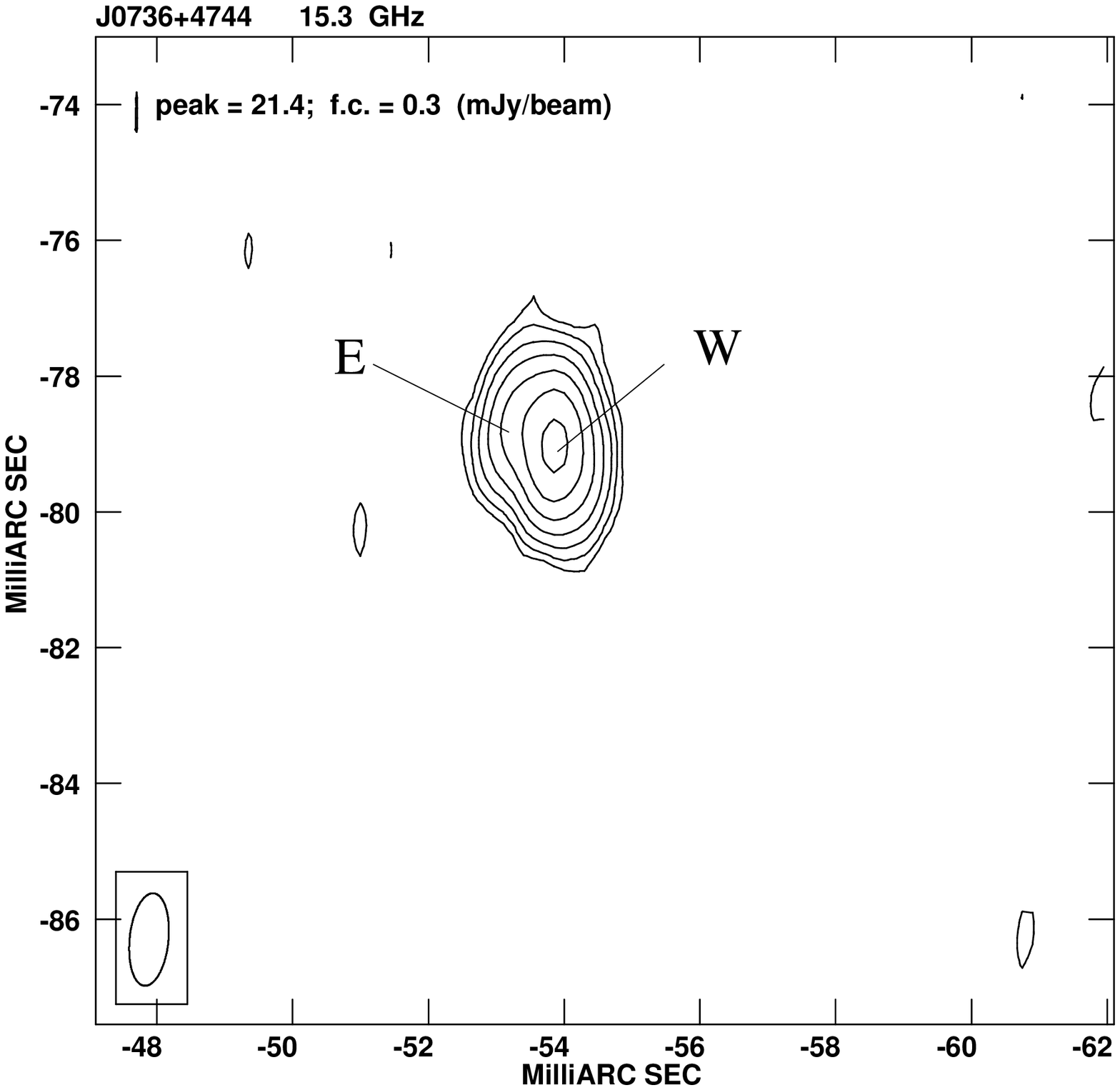}
\includegraphics{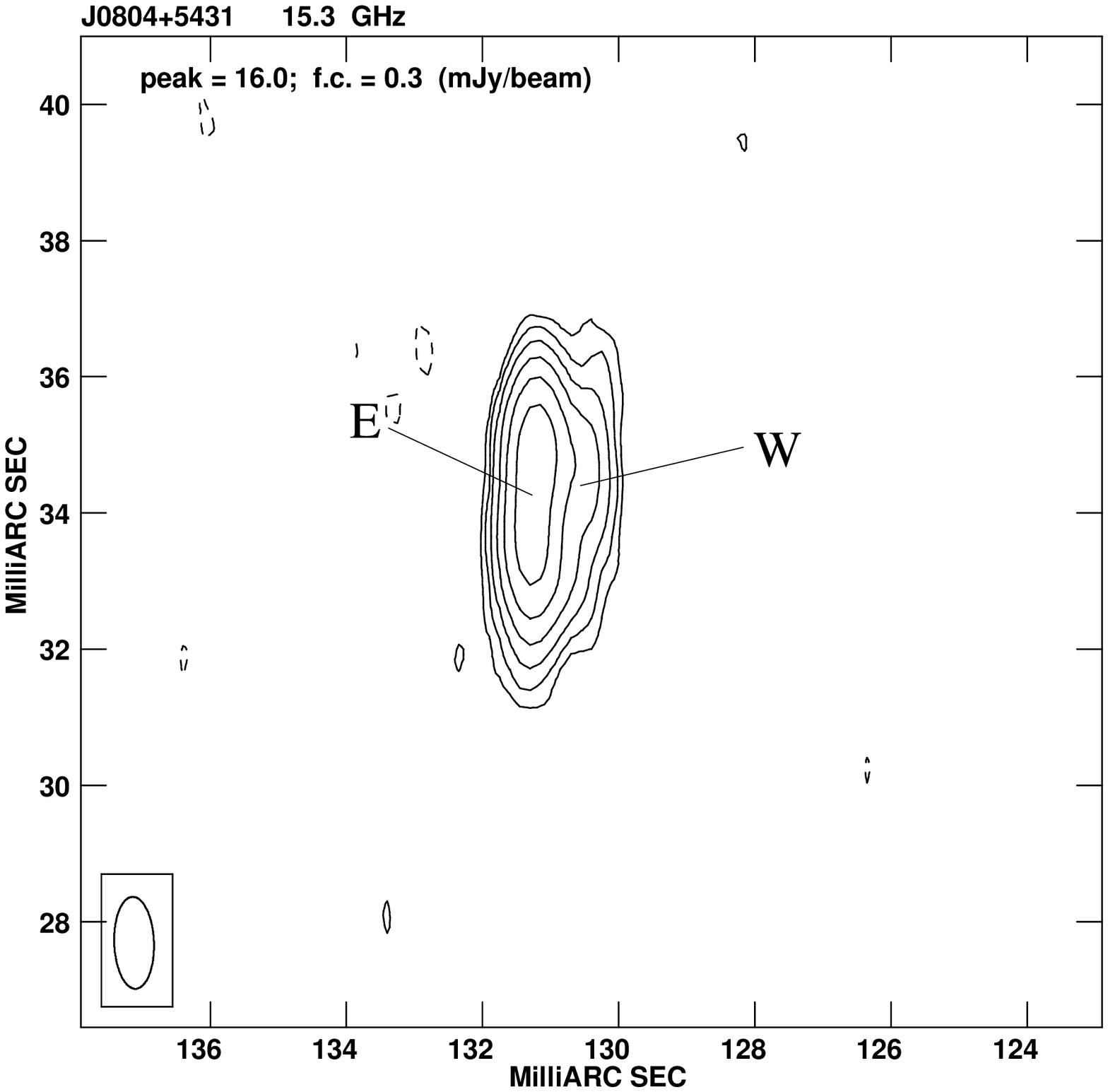}
\includegraphics{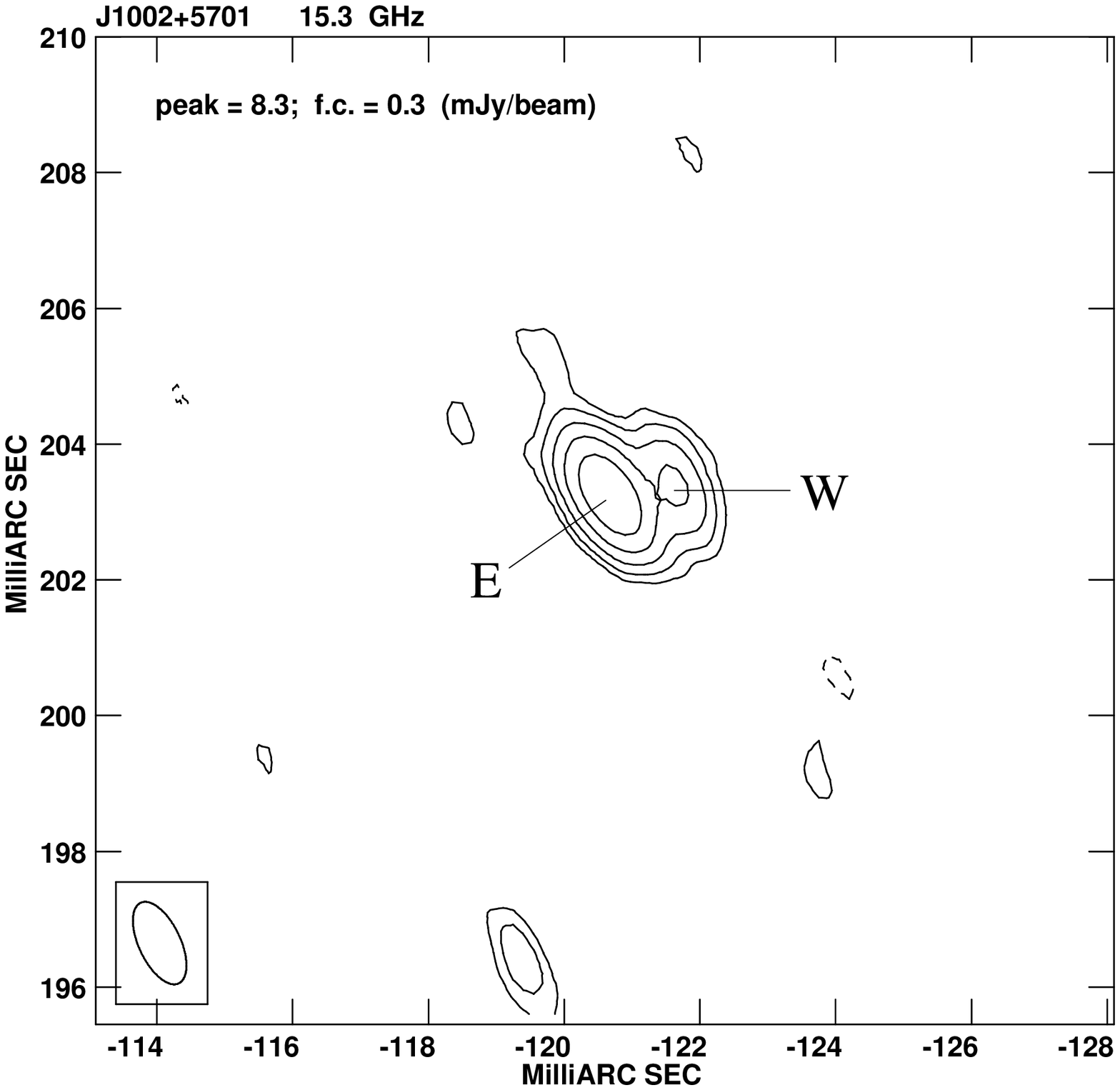}
\includegraphics{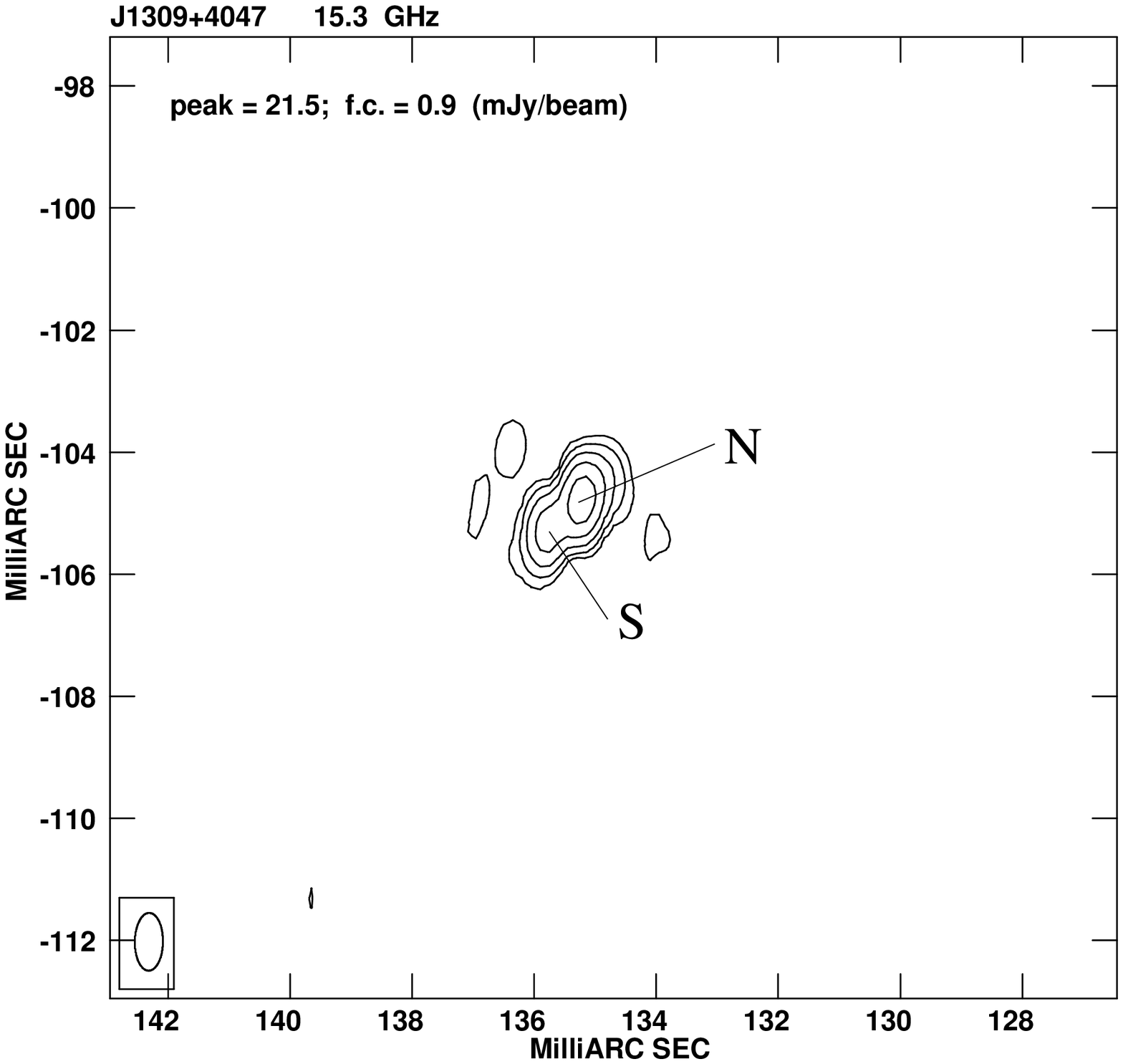}
\includegraphics{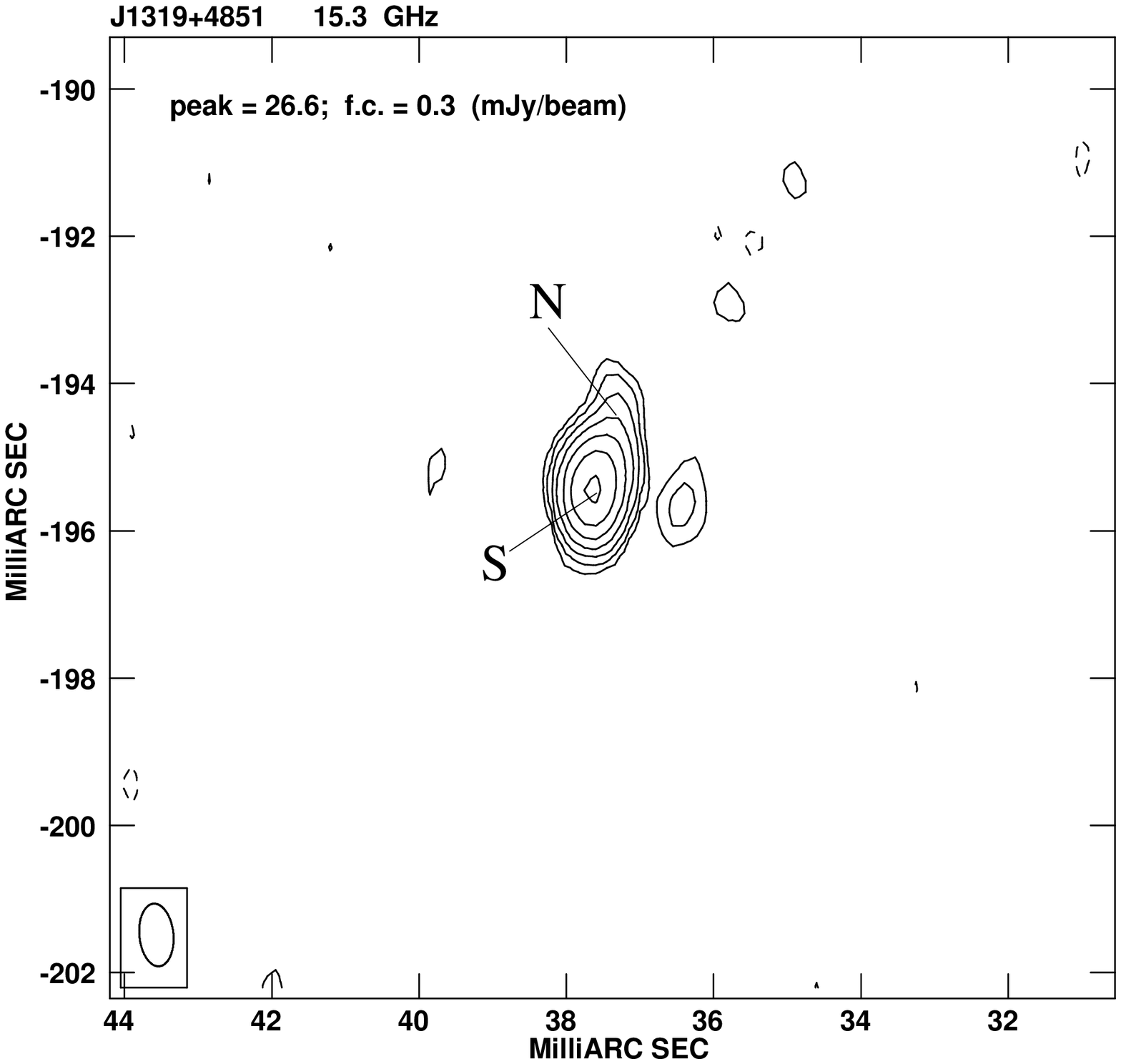}
\vspace{21.5cm}
\caption{VLBA images at 15.3 GHz of the sources with a marginally
  resolved structure. For each
  image we provide the observing frequency, peak flux density
  (mJy/beam), the first contour intensity which corresponds to three
  times the off-source noise level
  on the image plane. Contours increase by a
  factor of 2. The restoring beam is plotted on the bottom left
  corner.
Positions are relative to those used in the correlator.}
\label{mr}
\end{center}
\end{figure*}

\begin{figure*}
\begin{center}
\includegraphics{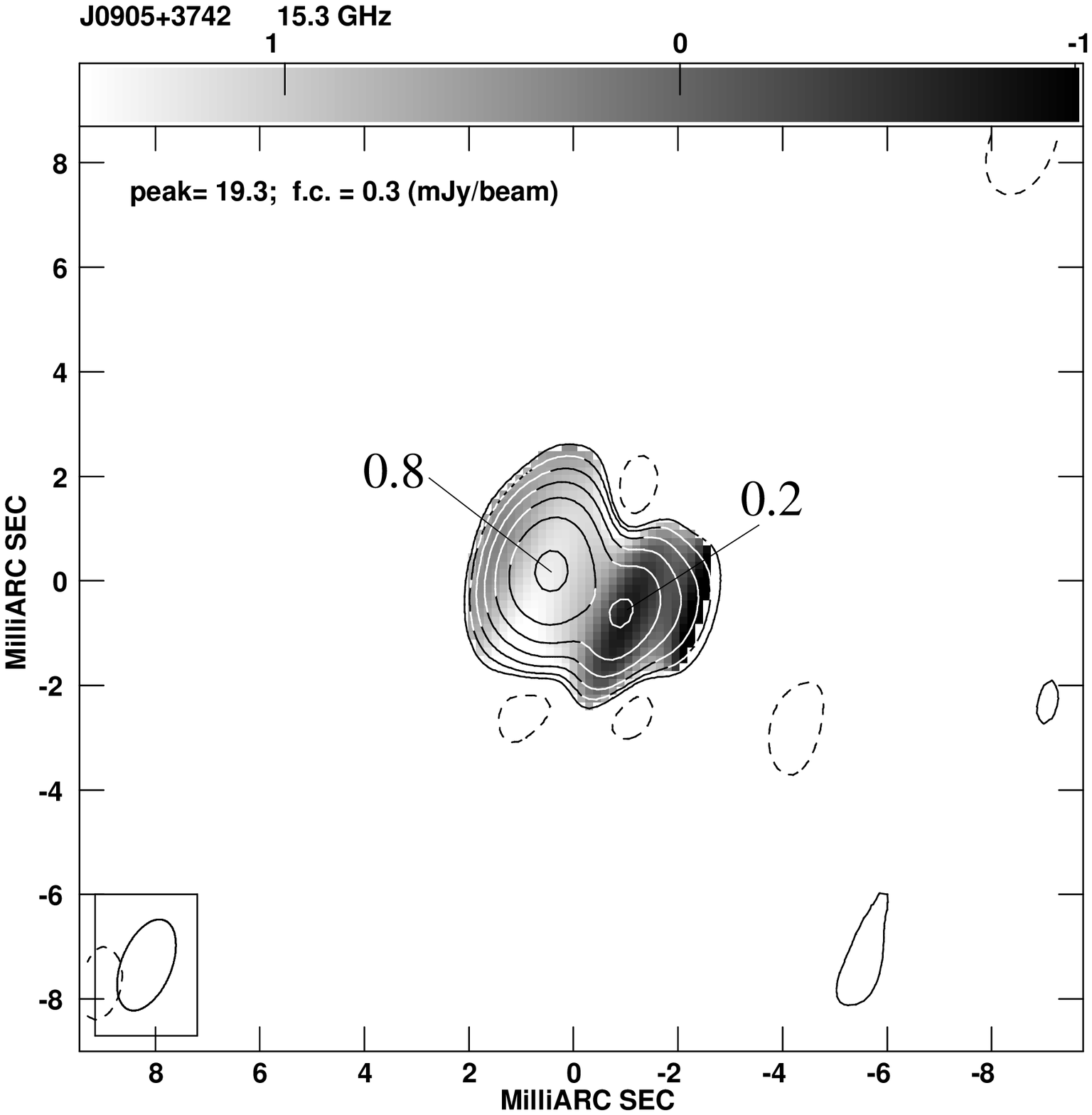}
\includegraphics{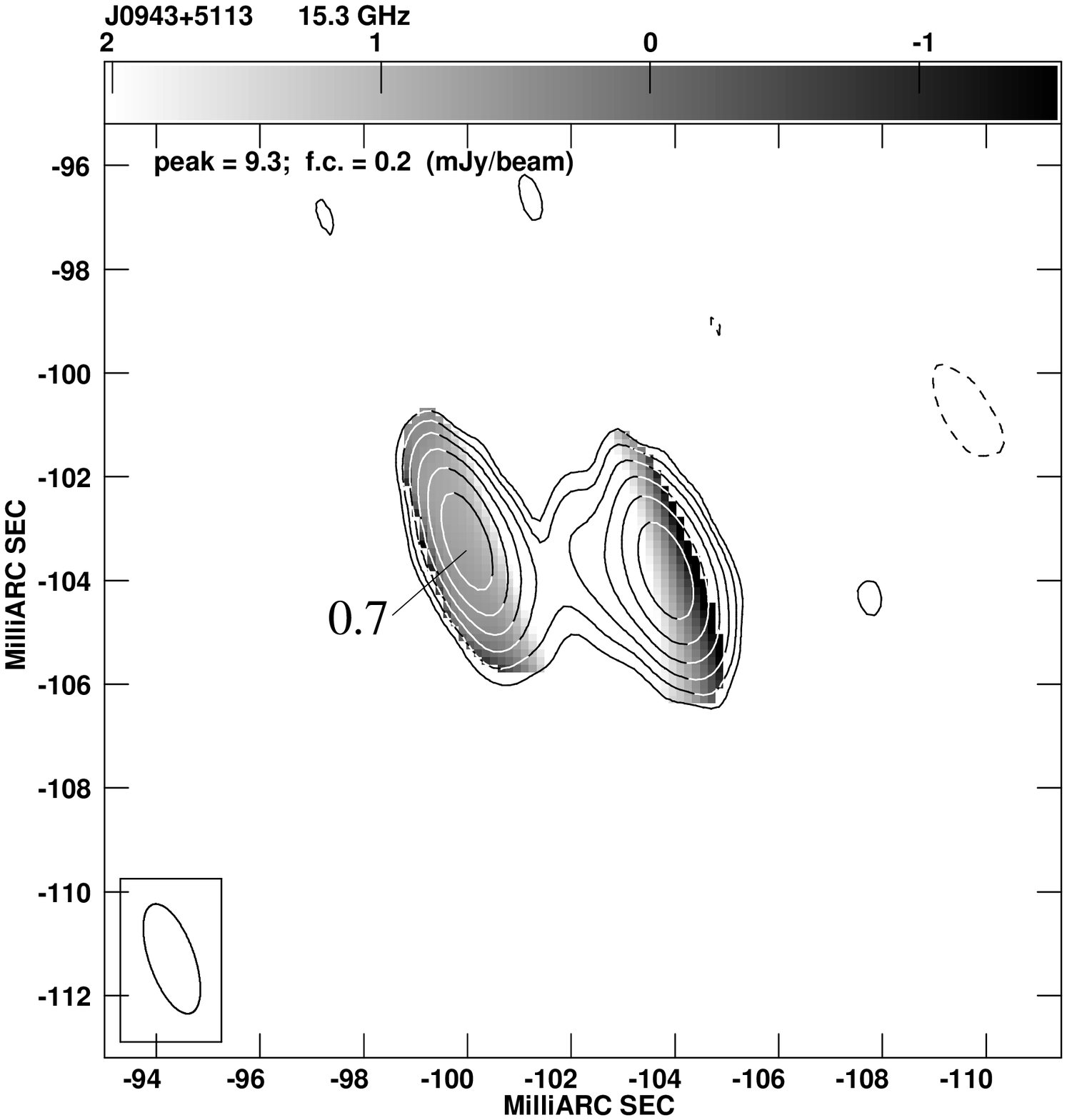}
\includegraphics{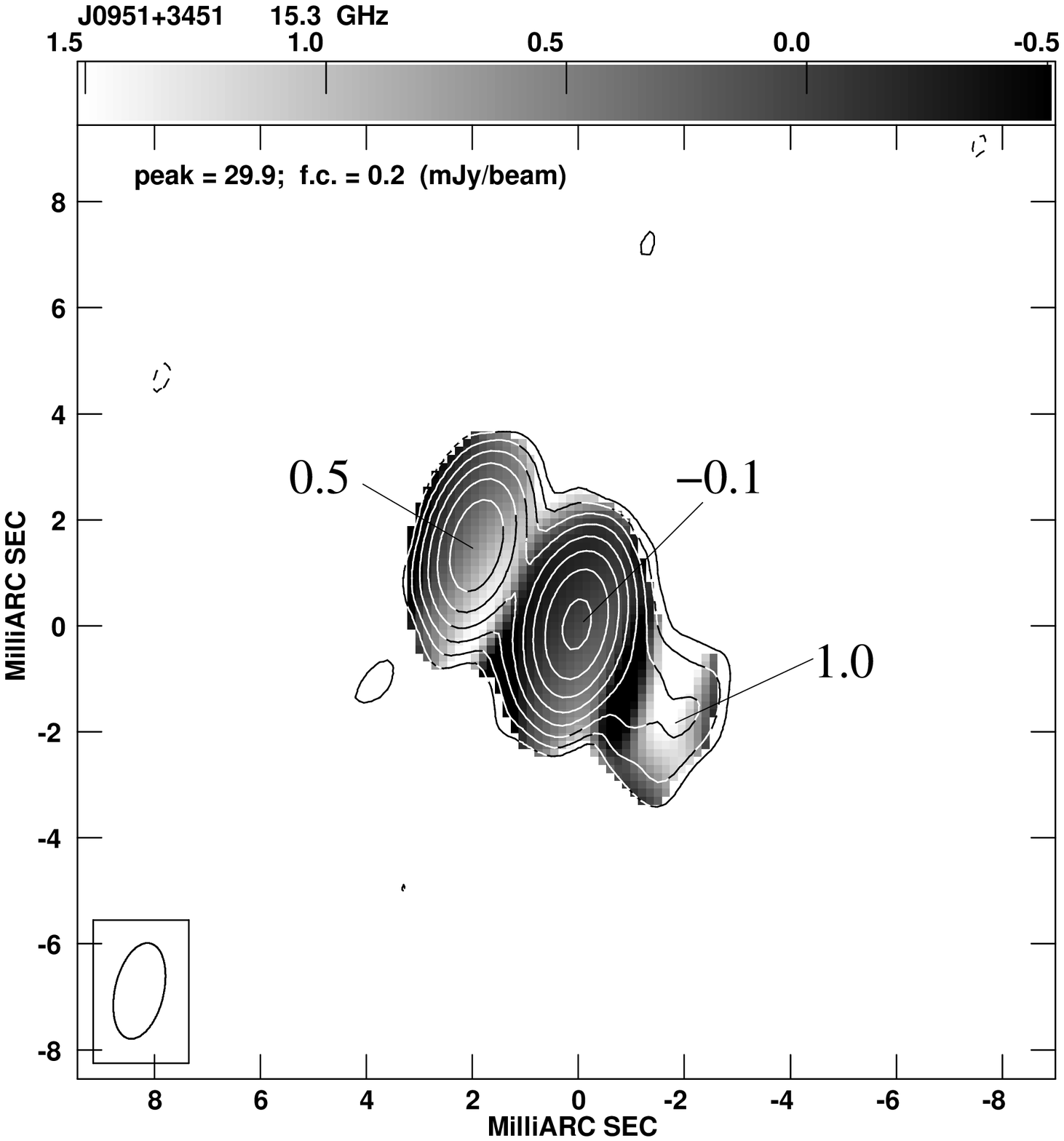}
\includegraphics{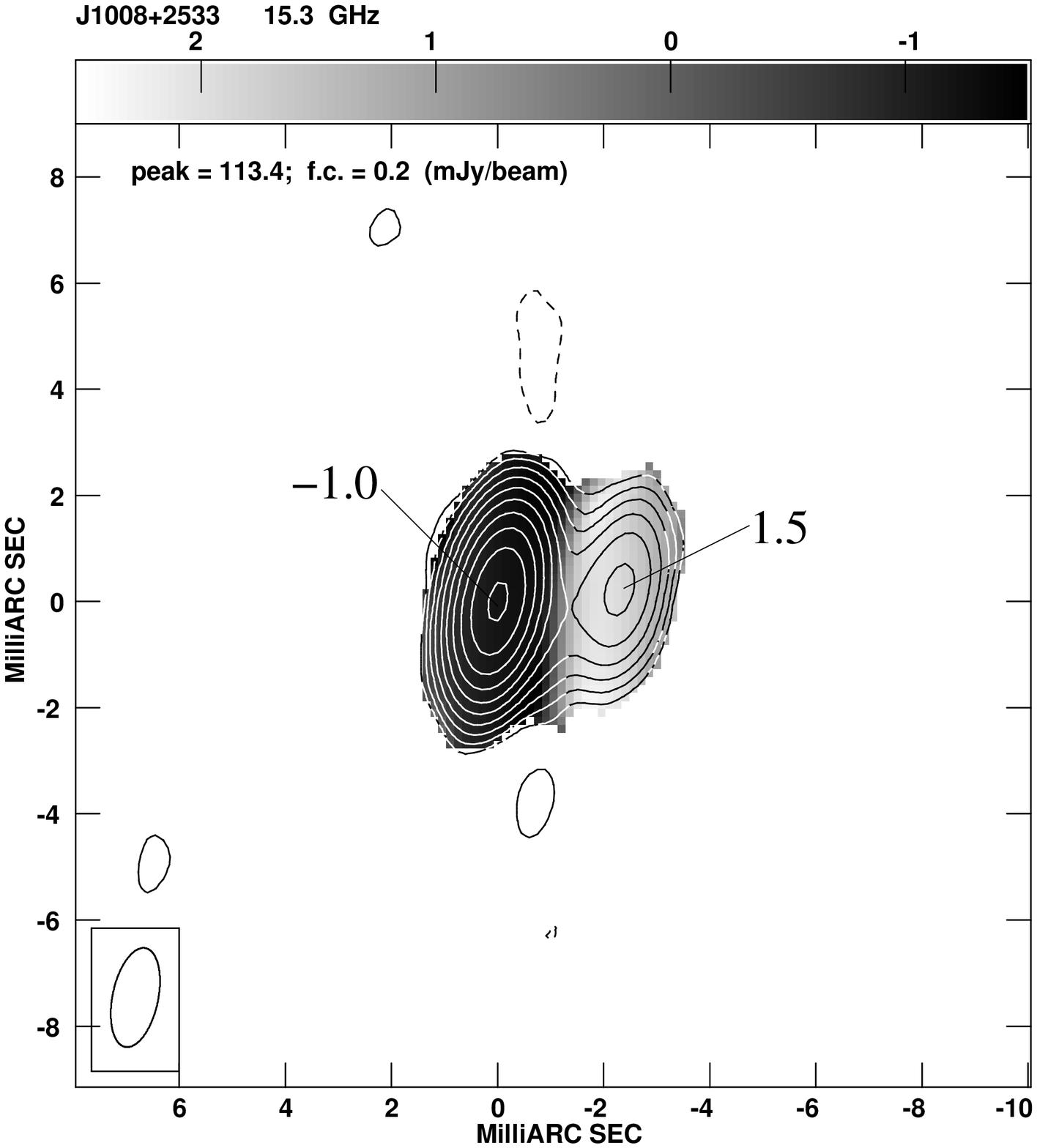}
\includegraphics{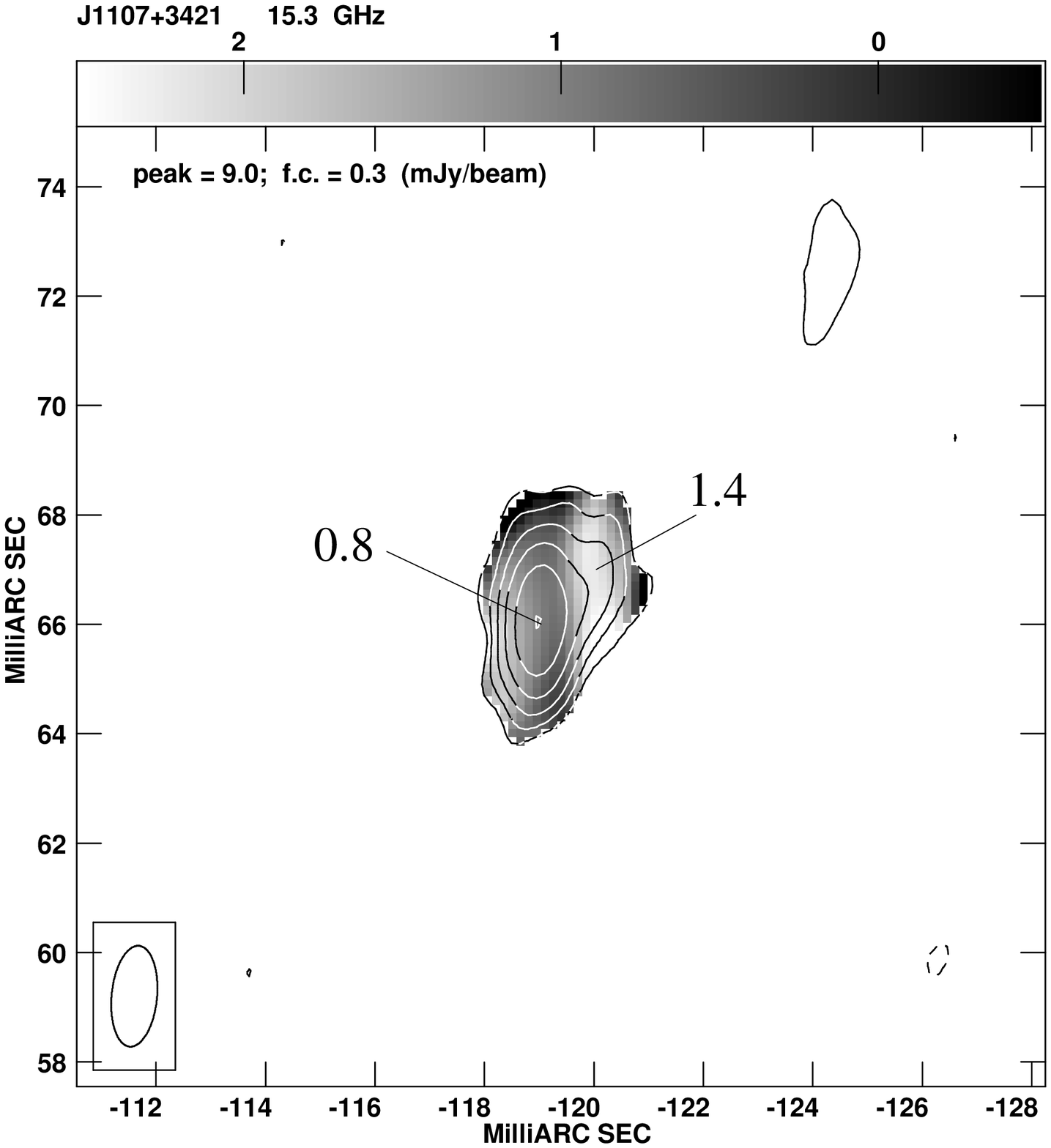}
\includegraphics{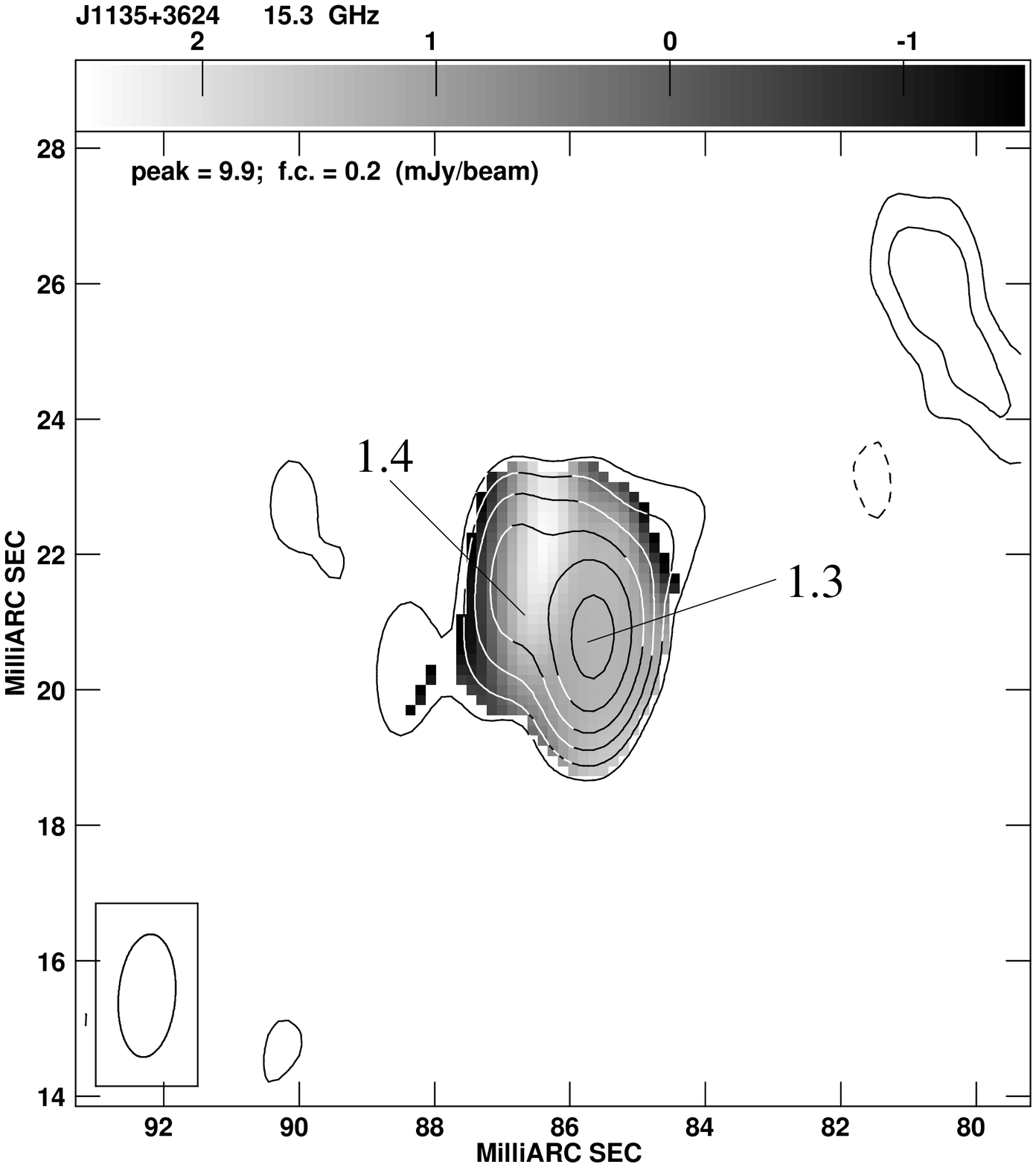}
\vspace{21.5cm}
\caption{Greyscale spectral-index images between 8.4 and 15.3 GHz of
  the sources with a resolved structure at both frequencies,
  superimposed on the low-resolution 15.3-GHz contour image 
convolved with the 8.4-GHz
  beam. For each
  panel we provide the observing frequency, peak flux density
  (mJy/beam), and the first contour intensity which corresponds to three
  times the off-source noise level 
on the image plane. Contours increase by a
  factor of 2. The restoring beam is plotted on the bottom left
  corner.
Positions are relative to those used in the correlator.}
\label{spix}
\end{center}
\end{figure*}

\subsection{Notes on individual sources}

In this section we discuss in detail the observational properties of
the sources with resolved morphology.
In general the spectral index distribution is uniform across the source
components, and the value computed 
by means of the total flux density of the whole component
measured from the full-resolution images at 8.4 and 15.3 GHz
(Table \ref{tab_comp}),
agrees with
the one directly measured on the spectral index image
(Fig. \ref{spix}). The use of the total flux density in computing the
spectral index may cause an artificial steepening 
due to the absence of the shortest spacing at the higher
frequency. Since our targets do not possess diffuse emission,
the estimated spectral indices should not be severely affected by the
different {\it uv}-coverages.
However, there are a few source components, remarked in the
following discussion, where a
non-artificial spectral index gradient is present. \\
Following \citet{mo06}, we consider CSO candidates (labelled CSO? in
Table \ref{sample}) those sources with
a spectral index between 8.4 and 15.3 GHz $>$0.5 (Table \ref{sample}), 
and a double/triple pc-scale
structure whose components have a steep spectral index (Table
\ref{tab_comp}). When the central core is present, we classify
  the source as a genuine CSO (labelled CSO in Table \ref{sample}). 
On the other hand,
we classify blazar objects (labelled BL in Table \ref{sample})
those sources with a spectral index $<$0.5
between 8.4 and 15.3 GHz (Table \ref{sample}) and/or with a
Core-Jet-like morphology, i.e. one-sided structure
dominated by a compact flat-spectrum
component (containing the core) from which a steep-spectrum jet-like
component emerges. When the source is unresolved and both the
  spectral index information and the variability behaviour is unclear,
we do not attempt any classification (the source is labelled ``?'' in
Table \ref{sample}.)\\

\subsubsection{Sources resolved at both 8.4 and 15.3 GHz}

Here we discuss the sources marked as ``Resolved'' in Table
\ref{sample}. The images of these sources are shown in Fig. \ref{cso}:
left and right panels present the VLBA image at 8.4 and 15.3 GHz
respectively. The spectral indices reported in the following
  discussion are those from Table \ref{tab_comp} and they were computed
  considering the total flux density of the whole component. In the
  case we consider the spectral index derived from the flux density
  peak on the spectral index images, we explicitly say that we are
  considering the local spectral index, and we make a reference to
  Fig. \ref{spix}.\\  

\noindent - J0905+3742: the radio structure of this source is
characterized by two compact and asymmetric components separated by
1.7 mas and position angle of about -130$^{\circ}$ (Fig. \ref{cso}). 
The flux density ratio S$_{\rm E}$/S$_{\rm W}$ is 3.6 and 3.0
at 8.4 and 15.3 GHz respectively. The eastern component has a steep
spectrum with $\alpha \sim 0.9$.
The spectral index distribution in the western 
component is more inhomogeneous, with an averaged spectral index
$\alpha \sim 0.6$, that can reach locally $\alpha \sim 0.2$  
(Fig. \ref{spix}). 
From its characteristics, we consider J0905+3742 a
CSO candidate. The flux density
obtained from our VLBA data is in good agreement with that derived from
earlier VLA observations at both frequencies. \\

\noindent - J0943+5113: this source has an asymmetric double structure
of 3.9 mas (21.5 pc at $z$=0.42) in size and position angle of about
85$^{\circ}$ (Fig. \ref{cso}). 
The flux density ratio S$_{\rm W}$/S$_{\rm E}$ is
3.6 and 1.2 at 8.4 and 15.3 GHz respectively, indicating a different
spectral index for the two components. Although the western
component is the brighter, it has a
very steep
spectral index ($\alpha = 2.1$), suggesting that relativistic particles
undergo severe energy losses not balanced by acceleration
processes. This value may be artificially steeper due to flux density
losses at 15.3 GHz, but it agrees with the spectral index
derived from the optically thin total spectral index 
\citep[$\alpha = 1.4$,][]{mo07},
indicating that the integrated source spectrum is dominated by the
western component. 
On the other hand, the spectrum of
the eastern component is flatter, with a local spectral
index $\alpha \sim 0.7$ (Fig. \ref{spix}). Diffuse emission
connecting the main components is detected at 8.4 GHz, whereas at 15.3
GHz it is visible in the low-resolution image only (Fig. \ref{spix}). 
On the basis of the morphological information, we classify J0943+5113 as a
CSO candidate. The flux density
obtained from our VLBA data is about 90\% of that derived from
VLA observations at both frequencies. This difference can be explained
either by some degree of variability, or by some missing flux density not
detected by the VLBA. No significant variability was found in earlier
VLA observations \citep{mo10b}.\\

\noindent - J0951+3451: this source shows a highly asymmetric triple
structure of 4.8 mas (20.7 pc at $z$=0.29) in size and position angle of
about 45$^{\circ}$ (Fig. \ref{cso}).
The central component has a flat spectrum ($\alpha \sim 0$) and likely
hosts the source core. The radio emission mainly comes from the central
and the eastern components while the western component represents only
7\% and 4\% of the total flux density at 8.4 and 15.3 GHz
respectively. The flux density ratio S$_{\rm E}$/S$_{\rm W}$ is 4.5
and 6.5 at 8.4 and 15.3 GHz, respectively. With a spectral index $\alpha
\sim0.5$, 
component E is likely a hotspot. Component W has a  
steeper spectrum ($\alpha \sim1.1$, Fig. \ref{spix}), and at 15.3 GHz
its structure is resolved into two aligned subcomponents.  
From its parsec-scale structure and the detection of the core,
we classify J0951+3451
a genuine CSO. Furthermore, the flux density
obtained from our VLBA data is in good agreement with that derived from
VLA observations at both frequencies. 
No significant variability was found in earlier
VLA observations \citep{mo10b}.\\

\noindent - J1008+2533: at 15.3 GHz the source displays a core-jet
structure with size of 2.6 mas (22.1 pc at $z$=1.96) 
and position angle of about -70$^{\circ}$,  
while at 8.4 GHz it appears as
a double with S$_{\rm E}$/S$_{\rm W}$ $\sim$ 2 (Fig. \ref{cso}). 
The eastern component has an inverted spectrum $\alpha \sim
-1.1$, while the jet-like feature has a steep spectral index $\alpha
\sim 1.8$. When imaged with the higher resolution of the 15.3-GHz data, 
an additional compact component, labelled C in Fig. \ref{cso}, 
is clearly visible at about 
1 mas (8.5 pc)
from the core, and it may be a knot of the jet. 
These characteristics suggest that J1008+2533 is a
blazar rather than a genuinely young object.    \\

\noindent - J1107+3421: this radio source has an asymmetric double
structure of 1.2 mas in size and position angle of about -50$^{\circ}$
(Fig. \ref{cso}). The flux density ratio S$_{\rm E}$/S$_{\rm W}$ is
1.9 and 2.1 at 8.4 and 15.3 GHz, respectively.
Component E has a spectral index $\alpha \sim
0.8$, while component W, which is also the fainter one, has a steeper
spectrum. On the basis of these morphological properties 
we classify J1107+3421 a CSO candidate. However, the flux density
obtained from our VLBA data is about 65\% of that derived from
earlier VLA observations at both frequencies. This difference can be explained
either by some degree of variability, or by some flux density not
detected by the VLBA. No significant variability was found in earlier
VLA observations \citep{mo10b}.\\

\noindent - J1135+3624: this radio source shows an asymmetric double
structure of 1.3 mas and position angle of about
40$^{\circ}$ (Fig. \ref{cso}). The flux density ratio S$_{\rm
  W}$/S$_{\rm E}$ is 2.1 and 1.8 at 8.4 and 15.3 GHz respectively.
Component W is the brighter one and it
has a compact structure at both the observing frequencies, while
component E is resolved at 15.3 GHz. Both components have steep
spectra, locally reaching a spectral index value of 1.3 and 1.5 for
component W and E, respectively (Fig. \ref{spix}). 
On the basis of these morphological
properties we classify J1135+3624 a CSO
candidate. The flux density
obtained from our VLBA data is in good agreement with that derived from
VLA observations at both frequencies.
No significant variability was found in previous
VLA observations \citep{mo10b}.\\

\subsubsection{Sources resolved at 15.3 GHz only}

Here we discuss the sources marked as ``Marginally Resolved'' in Table
\ref{sample}. For these sources only the VLBA image at 15.3 GHz is 
shown in Fig. \ref{mr}, since in the 8.4-GHz image 
they appear point-like. For these sources the only value of the 
spectral index is that
  computed from the total flux density measured on the full-resolution
images at 8.4 and 15.3 GHz. For this reason flux
density and spectral index variations, with respect to those
values reported in \citet{mo10b}, are not independent.\\

\noindent - J0736+4744: this source is marginally resolved into two
compact components separated by 0.8 mas in position
angle of about 70$^{\circ}$ (Fig. \ref{mr}). 
The spectral index computed considering the VLBA flux
density integrated on the whole source is $\alpha \sim 0.5$, slightly
flatter than the value derived from the VLA data \citep[$\alpha \sim
  0.8$][]{mo10b}. 
The flux densities
obtained from our VLBA data is in good agreement with that derived from
VLA observations at both frequencies. However, the lack of a
  secure morphological classification does not allow us to reliably
  constrain the nature of this source.\\

\noindent - J0804+5431: this source is marginally resolved into two
components separated by 0.9 mas (3.2 pc at $z$=0.22) in position
angle of about $-60^{\circ}$ (Fig. \ref{mr}). 
The spectral index computed considering the VLBA flux
density integrated on the whole source is $\alpha \sim 0.4$,
in agreement with the one 
derived from the VLA data \citep{mo10b}. 
The flux density
obtained from our VLBA data is about 80\% of that derived from
VLA observations at both frequencies. This difference can be explained
either by some degree of variability, or by some flux density not
detected by the VLBA. No significant variability was found in earlier
VLA observations \citep{mo10b}.
However, the lack of a
  secure morphological classification does not allow us to reliably
  constrain the nature of this source.\\

\noindent - J1002+6701: this source is resolved into two
components separated by 1.5 mas in position
angle of about -70$^{\circ}$ (Fig. \ref{mr}). 
The spectral index computed considering the VLBA flux
density integrated on the whole source is very steep ($\alpha \sim
2.3$). Although such a steep value may be due to missing flux density at the
higher frequency, it must be noted that the value derived from
VLA data is also very steep \citep[$\alpha \sim 1.8$,][]{mo10b}. 
On the basis of this steep spectral index 
we consider J1002+6701 a CSO candidate. However, the flux density
obtained from our VLBA data is about 75\% of that derived from
VLA observations at both frequencies. This difference can be explained
either by some degree of variability, or by some flux density not
detected by the VLBA. No significant variability was found in earlier
VLA observations \citep{mo10b}.\\

\noindent - J1309+4047: this source is resolved into two
components separated by 0.8 mas (6.3 pc at $z$=2.91) in position
angle of about 140$^{\circ}$ (Fig. \ref{mr}). 
The spectral index computed considering the VLBA flux
density integrated on the whole source is steep ($\alpha \sim
1.4$). Such a steep value may be due to missing flux density at the
higher frequency. The spectral index derived from
VLA data is $\alpha \sim 0.9$ \citep{mo10b}. 
On the basis of this steep spectral index 
we still consider J1309+4047 a CSO candidate. However, the flux density
obtained from our VLBA data is about 90\% and 65\% of that derived from
VLA observations at 8.4 and 15.3 GHz, respectively. 
This difference can be explained
either by some degree of variability, or by some flux density not
detected by the VLBA. No significant variability was found in earlier
VLA observations \citep{mo10b}.\\ 

\noindent - J1319+4851: this source is marginally resolved into two
components separated by 0.6 mas in position
angle of about 30$^{\circ}$ (Fig. \ref{mr}).
The spectral index computed considering the VLBA flux
density integrated on the whole source is inverted ($\alpha \sim
-0.2$), while from previous VLA observations, 
between these frequencies the spectrum was
already optically thin with $\alpha \sim 0.5$ \citep{mo10b}. 
On the basis of these spectral characteristics 
we consider J1319+4851 a blazar object rather than a genuine CSO. 
The flux density
obtained from our VLBA data is about 85\% of that derived from
VLA observations at 8.4 GHz, while at 15.3 GHz the VLBA and VLA flux
densities are in good agreement. 
No significant variability has been previously detected
from the VLA monitoring campaign \citep{mo10b}.\\

\section{Discussion}

Multifrequency observations with parsec-scale resolution 
are necessary to characterize the
morphology of compact radio sources. The addition of the information
on the spectral
index distribution in the optically-thin part of the 
spectrum enables a suitable
classification of each source component as hotspot, jet, core, or lobe,
providing a crucial tool to determine whether the source is either genuinely
young or a blazar. The optically-thin steep integrated spectral index
guarantees that the most luminous components are likely hotspots
rather than cores, at least at the two frequencies considered here. In
most cases, the flux density measured in our VLBA images agrees with
that detected by VLA observations a few years
earlier. This means that the majority of the sources 
are not significantly variable, contrary to that
expected in the case of blazars. A difference between VLBA and VLA
detected
flux densities may be related to some extended low-surface brightness
emission due to the insufficient
dynamic range of our observations.

\subsection{The source morphology}

From our VLBA observations we could resolve the source structure of 11
sources ($\sim$65\%) at least at 15.3 GHz. Among them,
nine objects ($\sim$82\%) show a double
structure (J0736+4744, J0804+5431, J0905+3742, J0943+5113, J1002+5751,
J1107+3421, J1135+3624, J1309+4047, and J1319+4851), while two ($\sim$18\%) 
are resolved into three well-defined components (J0951+3451 and J1008+2533). 
For the six sources
whose structure is resolved also in the 8.4-GHz data, we could
classify the nature of each sub-component by means of the information
on the spectral index. Compact components with spectral indices
between 0.5 and 0.8, likely hotspots, are found in four sources
(the eastern component of the sources J0905+3642, J0943+5113, J0951+3451, 
and J1107+3421). Despite its nearly flat spectral index ($\alpha = 0.2$),
we consider component W of J0905+3742 a very compact hotspot,
rather than the true source core, 
with the spectral peak occurring at a frequency close to 8.4 GHz 
which causes an artificial flattening of the spectrum, as in the
case of the very compact hotspots found in a few HFPs from the
bright HFP sample \citep[e.g. J1335+5844 and
  J1735+5049,][]{mo06}.\\
The core component is unambiguously detected in
two sources: the galaxy J0951+3451 and the quasar J1008+2533. In the
former object the core is the central component, almost midway between
the source hotspots/lobes. This source resembles the bright HFP quasar
J0650+6001, where the asymmetric properties are well explained by
Doppler boosting effects \citep{mo10a}.  
However, J0951+3451 is associated with a galaxy where projection
effects should not play a relevant role, as also suggested by the lack
of both flux density and spectral variability in this object \citep{mo10b}.
The different characteristics 
shown by the two external components might also
  be related to an inhomogeneous external medium surrounding the radio
  source. The interaction of one jet with a denser medium, like a cloud, 
  may slow down the
  jet expansion, decreasing the adiabatic losses 
  and producing an
  enhancement of the radio luminosity due to particle re-acceleration
  and synchrotron losses, 
as also found in many young CSS
sources \citep[see e.g.][]{labiano06,mo07}. A similar result may be
obtained when one jet has, for some reasons, an intrinsically 
higher internal pressure than the
other.
Another possible explanation is that component C consists of
two components, i.e. the true core and a hotspot, 
as suggested by its resolved structure in the 15.3 GHz image. In this
case J0951+3451 may be interpreted as an asymmetric
source, where the core is closer to the brighter hotspot. On the other
hand, component W, with its steep spectral index ($\alpha \sim 1.0$)
may represent a lobe-like feature where no particle acceleration 
is currently taking place. \\
In J1008+2533 the core component is the
brightest one and it is located at its easternmost edge, from
which a steep-spectrum jet emerges to the West.
This source resembles the
bright HFP quasar J2136+0041 \citep{mo06}. Although both sources possess some
amount of flux density variability, J2136+0041 preserves its convex
spectral shape \citep[e.g.][]{mo07}, 
whereas in J1008+2533 the radio spectrum during the
last two observing epochs shows a flattening between 5 and 43 GHz,
interpreted as the composition of two different spectra: convex below
8.4 GHz and inverted at higher frequencies \citep{mo10b}.  \\
From this analysis we found that in five
sources (J0905+3742, J0943+5113, J0951+3451, J1107+3421, and
J1135+3524) the radio emission has a double/triple structure
typical of young objects, 
while J1008+2533
displays a core-jet morphology, typical of blazars.\\
For the five sources marginally resolved at 15.3 GHz only, and for the
six sources unresolved at both frequencies, the lack of
information on the spectral index distribution 
does not allow us to classify their
sub-components. However, if we consider the integrated spectral index 
we find that in two objects (J1241+3844 and J1319+4851) the spectrum 
turned out to be flat or inverted between 8.4 and 15.3 GHz, instead of
steep as derived from the multi-epoch VLA campaign
\citep{mo10b}. This strongly indicates the blazar nature of these
two sources, and for this reason we drop these objects from the
sample of genuinely young radio source candidates.\\ 

\subsection{Flux density variability}

In general, the flux densities measured in our VLBA images, and
reported in Table \ref{sample}, is a significant
fraction, i.e. more than 80\%, 
of those detected by VLA observations a few years
earlier \citep{cstan09,mo10b}. 
This suggests that the majority of the observed sources (about 60\%)
are not significantly variable. 
In the remaining seven sources, 
the decreased flux densities
detected by the VLBA with respect to those obtained from VLA observations
may be related to some extended low-surface brightness
emission that cannot be accounted for due to the insufficient
dynamic range of our observations. An example may be the radio galaxy
J0943+5113, where diffuse emission connecting the main components is
detected at 8.4 GHz, while it is resolved out at 15.3 GHz. 
In five sources (i.e. J0955+3335, J1002+5701,
J1107+3421, J1436+4820, and J1613+4223) the flux density is
continuously decreasing with time, when the VLA flux densities at 
the various epochs are compared. 
This behaviour may be explained by effective energy
losses unbalanced by the acceleration of fresh particles,
thus related to the evolution of a genuinely young radio source,
rather than by variability related to boosting effects, like in
blazars. However, in the case of J0955+3335 in addition to a decrease
of the flux density, there is also a change in the spectral shape,
which becomes much flatter, $\alpha \sim 0.4$, than in earlier VLA
observations where the spectral index between 8.4 and 15.3 GHz was
0.7. Such a flattening, occurring in the optically-thin part of the
spectrum, suggests that the variability in J0955+3335 is likely related
to boosting effects as it happens in blazars.\\  
The sources J1008+2533 and J1319+4851 have similar VLBA and VLA flux
densities at 8.4 GHz, while at 15.3 GHz our VLBA observations show a
much higher flux density. This behaviour might be related to an increase of
the core activity.\\
In summary, if we take into account all the pieces of information 
on the structure, spectral
index and variability, we conclude that 6 objects (35\%), labeled CSO? in
Table \ref{sample}, are to be
considered CSO candidates, one object (6\%) is a confirmed CSO, 
4 objects (24\%), labeled BL in Table
\ref{sample}, are part of the blazar population, while for the
remaining 6 objects (35\%) there is not enough information 
to reliably constrain the nature of their radio emission.\\

\subsection{Physical properties}

\begin{figure}
\begin{center}
\includegraphics{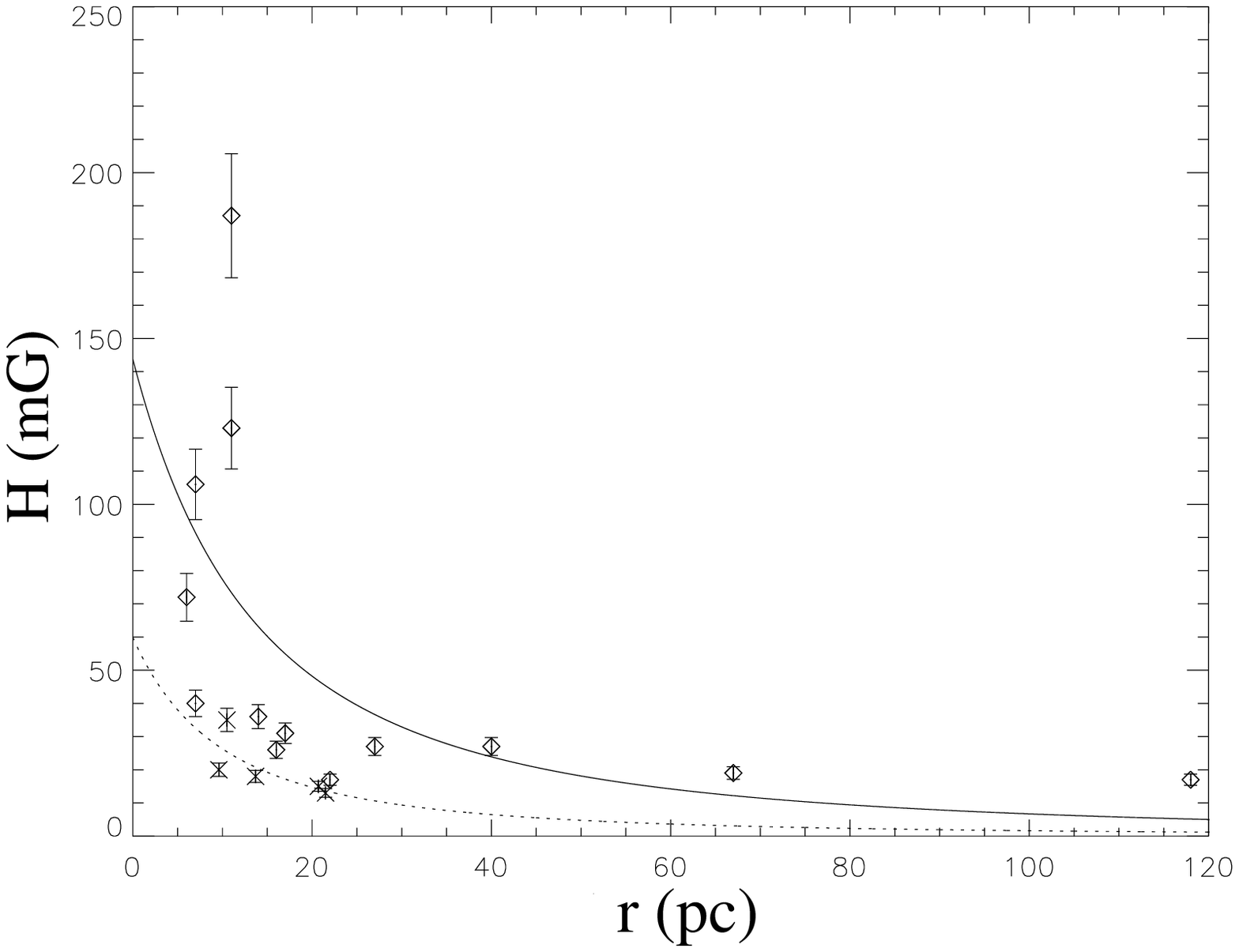}
\vspace{8.0cm}
\caption{Equipartition magnetic field $H$
versus the linear size $r$ of the genuine
  CSOs from the faint ({\it X symbols}), and the bright ($diamonds$)
samples. The lines represent the best fit to the model given in
  Equation \ref{adiabatic} for the genuine CSOs and CSO candidates
  from the bright ({\it
  solid line}), and the faint ({\it dotted line}) sample, 
as described in Section 4.3. } 
\label{h_lls}
\end{center}
\end{figure}

\begin{figure}
\begin{center}
\includegraphics{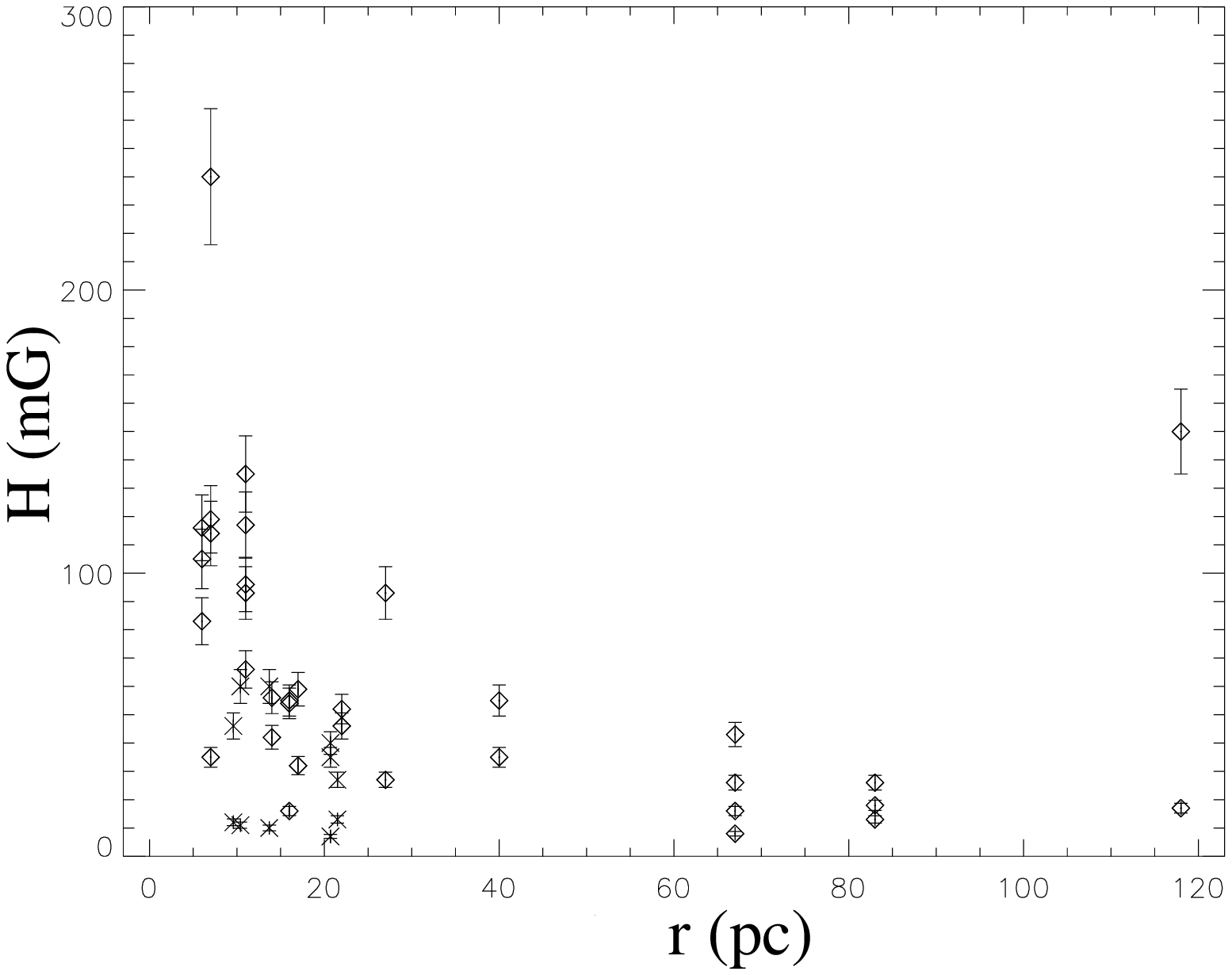}
\vspace{8.0cm}
\caption{Single component equipartition magnetic field $H$ versus the source
  linear size $r$ of the components
  of the genuine
  CSOs and CSO candidates from the faint ({\it X symbols}), 
and the bright ($diamonds$)
samples.}
\label{comp_lls}
\end{center}
\end{figure}

The knowledge of physical conditions in young radio sources is
  important for defining the framework of models describing radio
  source evolution.
To draw a more comprehensive picture of the individual source evolution,
we compare the physical parameters of these small and faint sources
to those derived in other samples (both bright and faint) of
HFP/GPS/CSS \citep[e.g.][]{dd00,fanti90,fanti01}. \\
To adopt the same approach used in earlier works on CSS/GPS/HFP
samples, we compute the physical parameters assuming that the sources
are in the 
minimum energy condition, which corresponds to equipartition of energy
between magnetic field and radiating particles
\citep{pacho70}. 
Evidence in favour of this hypothesis has been found by \citet{mo08b}
by comparing the equipartition magnetic field with that computed on
the basis of observed quantities like peak flux density, peak frequency, and
linear size in self-absorbed components in a number of HFPs from the
bright sample. In the present work we cannot perform such a
computation since the two-frequency observations presented here are
not adequate to estimate spectral 
peak values (frequency and flux density) for
the individual components.\\
We computed the minimum energy density, minimum pressure and the
equipartition magnetic field by means of the standard formulae
\citep{pacho70}, assuming equal energy between protons and
electrons ($k$=1), and an average optically-thin
spectral index of 0.7. \\
The minimum energy density $u_{\rm min}$ is derived from:

\begin{equation}
u_{\rm min} = 7 \times 10^{-24} \left( \frac{L}{V} \right)^{4/7}
(1+k)^{4/7}\,\,\,\,{\rm erg\,cm^{-3}}
\label{energy}
\end{equation}

\noindent where $L$ is the luminosity at 8.4 GHz in W~Hz$^{-1}$, and $V$ the
volume in kpc$^{3}$, computed assuming an ellipsoidal geometry and a
filling factor of unity (i.e. the source is fully and homogeneously
filled with the relativistic plasma):

\begin{equation} 
V = \frac{\pi}{6}d_{\rm min}^{2} d_{\rm max}
\label{volume}
\end{equation}

\noindent where $d_{\rm max}$ and $d_{\rm min}$ are the component major and
minor axes, in pc. The equipartition magnetic field is computed by:  

\begin{equation}
H_{\rm eq} = \sqrt{ \frac{24}{7} \pi u_{\rm min}}
\label{campo}
\end{equation}

\noindent while the minimum pressure $p_{\rm min}$ is $\sim$(1/3)$u_{\rm
min}$.\\
From these equations we find that the typical values for the
components of the faint HFP sample are:
$u_{\rm min} = 10^{-4}-10^{-5}$ erg/cm$^{3}$, $p_{\rm min} = 10^{-4} -
10^{-6}$ dyne/cm$^{2}$, and $H_{\rm eq}$ ranges between 7 and $\sim$60 mG. 
Errors have been estimated by means
  of the propagation error theory and they are about 10\% for $H_{\rm eq}$. 
However, we must note that the uncertainties in the
  minimum energy parameters are strictly related to the assumptions of
  the value of
  both $k$ and the filling factor, which may vary by several orders of 
magnitude.\\
The magnetic field values obtained are smaller than those 
usually found in the components of bright HFPs \citep{mo06},
but they are in good agreement with those derived 
in the compact hotspots of CSS/GPS
sources with similar luminosity \citep{dd02}, while in the lobes of
CSS/GPS sources such
values are about 1-2 order of magnitude lower
\citep[e.g.][]{fanti95}.
This result indicates
that in the HFP sources  
the radio emission mainly comes from compact regions rather than from
extended structures like lobes.  
In fact, the radiative lifetimes of
relativistic electrons in such strong magnetic fields are very short
and they are not expected to form a detectable back-flow lobe.\\
In Fig. \ref{h_lls}, for the HFPs reported in Table \ref{tab_comp}
with a double/triple
structure consistent with a young radio source, 
we show the magnetic field averaged over the
whole source, obtained by considering the whole elliptical volume of
the entire source, computed by means of Eq. \ref{volume}, as filled by
plasma (represented with X symbols in Fig. \ref{h_lls}). 
As a comparison, the magnetic field
computed in the same way for the sources from the bright HFP sample
\citep{mo06} is shown as diamonds.
This plot shows that the faint HFPs in this study have linear sizes
comparable to those of the more compact bright HFPs but have magnetic
fields like those in the most extended bright HFPs. The most
compact (i.e. with sizes similar to those found in faint HFPs) and
bright HFPs have instead larger fields.
As expected, Fig. \ref{h_lls} shows that the higher magnetic
fields are found in the most compact and brightest objects. \\
To investigate a possible evolution of the magnetic field as the
source expands, we investigate its relation with the linear size $r$
assuming pure adiabatic expansion. First of all, we assume that the
radio emission is due to a homogeneous component that is adiabatically
expanding at a constant rate:

\begin{equation}
r = r_{0} \left( \frac{t_{0} + \Delta t}{t_{0}} \right)
\label{r_ad}
\end{equation}
 
\noindent where $r_{0}$ is the source size at time $t_{0}$ and $r$
the size at the time $t_{0} + \Delta t$.\\
We also assume that the magnetic field
is frozen into the plasma:

\begin{equation}
H = H_{0} \left( \frac{t_{0}}{t_{0} + \Delta t} \right)^{2}
\label{h_ad}
\end{equation}

\noindent where $H_{0}$ is the magnetic field at the time $t_{0}$ and
$H$ the magnetic field at the time $t_{0} + \Delta t$.\\
To test the case of pure adiabatic expansion, we fit 
the magnetic fields and the linear sizes of the sources from the 
bright and faint sample separately, 
with the function:\\

\begin{equation}
H(t)=H_{0} \left( \frac{r_{0}}{r_{0}+r(t)} \right)^{2}
\label{adiabatic}
\end{equation}

\noindent where $H_{0}$ and $r_{0}$ are the magnetic field and size 
normalization, respectively, for the faint and bright samples
separately. From Fig. \ref{h_lls} it is clear that
the fit does not follow the expected magnetic field decrease as the
source size increases, providing a reduced Chi-squared of 202 and
51 for the bright and faint sample, respectively. We must note
that Eq. \ref{adiabatic} represents a magnetic field that is frozen in
a homogeneous component that is adiabatically expanding at a constant
rate, without the injection/acceleration of new relativistic particles
and no radiative losses. However, in young radio sources radiative
losses play a major role given the high values of the magnetic
field. Radiative losses would decrease the flux density mainly in the
optically-thin part of the spectrum, thus underestimating the magnetic
field. On the other hand, 
the injection/acceleration of relativistic particles is still taking
place, as proved by the detection of the source core and
hotspots. The continuous refueling of relativistic particles, 
and thus energy and magnetic field, 
implies that the initial conditions we considered for a frozen
magnetic field are not suitable.
Furthermore, the source is not likely evolving as a
homogeneous component. 
Therefore, it is clear that the equipartition fields we determine as a
function of the source size (i.e. the source age) need a more complex
framework. The magnetic field can depend on the fate of
individual components of each radio source. To study this aspect, in
Fig. \ref{comp_lls} we plot the equipartition magnetic field, computed following Eq. \ref{campo} and listed in Table \ref{tab_comp}, of all
the components of each non-blazar object with resolved structure
(i.e. at least two components per
source) versus the source linear size. From this plot it is rather
clear that the field intensities found in the various components of
the same object can vary up to an order of magnitude. Such differences
may arise from asymmetries in the source propagation, for example
when the two sides experience a different environment.\\
All these pieces of evidence indicate that simple self-similar
evolution models are not adequate to describe the radio source growth,
as also pointed out by \citet{bicknell07}, \citet{bicknell03} and 
\citet{fanti03}.
To
reproduce the evolution of the magnetic field, and thus of the radio
source itself, a more complicated model taking into account the various
physical and observational properties, as well as the ambient
conditions must be considered.\\

\section{Summary}

We presented VLBA observations at 8.4 and 15.3 GHz of 17 sources from
the ``faint'' HFP sample \citep{cstan09}. The sources were selected on
the basis of their peak frequency occurring below 8.4 GHz, in order to
study their properties in the optically-thin part of the
spectrum. We found that 11 objects are resolved into two/three
components, while 6 are unresolved even at 15.3 GHz. When we complement
the information on the structure with that on the spectral index
distribution 
we found that in 7 of the resolved objects
the radio emission is likely dominated by hotspots, as in genuine
young radio sources, whereas 2 objects display a core-jet structure, typical of
the blazar population. In the case of the unresolved sources, although
nothing could be said on their structure, the availability of the two
frequencies allowed us to find 2 sources 
whose spectra turned out to
be flat/inverted between 8.4 and 15.3 GHz, indicating a change in the
spectral index with respect to that derived from previous VLA
observations \citep{mo10b}, and thereby implying strong variability
in these sources. The component magnetic fields 
are similar to the values in the hotspots of young sources with
larger sizes, but smaller than those found in
the ``bright'' HFPs. 
Such high magnetic fields cause severe radiative losses,
precluding the formation of extended lobe structures. \\

\begin{table}
\caption{Accurate position of the sources observed with the
  phase-referencing technique. The uncertainty on the source
    position is 0.5 mas (see Section 3).}
\begin{center}
\begin{tabular}{cccc}
\hline
Source& RA (J2000) & Dec (J2000) &Cal. source\\
\hline
J0736+4744& 07:36:01.0496& 47:44:23.991&J0742+4900\\
J0804+5431& 08:04:59.2531& 54:31:57.804&J0742+4900\\
J0943+5113& 09:43:51.8174& 51:13:22.537&J0929+5013\\
J1002+5701& 10:02:41.6662& 57:01:11.483&J0921+6215\\
J1107+3421& 11:07:34.3384& 34:21:18.596&J1130+3815\\
J1135+3624& 11:35:52.2921& 36:24:22.011&J1130+3815\\
J1241+3844& 12:41:43.1334& 38:44:04.271&J1242+3751\\
J1309+4047& 13:09:41.5089& 40:47:57.235&J1329+5009\\
J1319+4851& 13:19:30.2948& 48:51:03.194&J1329+5009\\
J1420+2704& 14:20:51.4880& 27:04:27.045&J1407+2827\\
J1436+4820& 14:36:18.9097& 48:20:41.135&J1407+2827\\
\hline
\end{tabular}
\end{center}
\label{astrometry}
\end{table}

\section*{Acknowledgments}
We thank the referee Jonathan Marr for carefully reading
the manuscript.
The VLBA is operated by the US National Radio Astronomy Observatory which
is a facility of the National
Science Foundation operated under cooperative agreement by Associated
Universities, Inc. This work has made use of the NASA/IPAC
Extragalactic Database NED which is operated by the JPL, Californian
Institute of Technology, under contract with the National Aeronautics
and Space Administration.

\end{document}